\documentclass[aps,twocolumn,pra,superscriptaddress,amsmath,showpacs,tightenlines,pdflatex,longbibliography]{revtex4-1}
\usepackage{amssymb}
\usepackage{amsmath}
\usepackage{dcolumn}
\usepackage{graphicx}
\usepackage{mathrsfs}
\usepackage{appendix}
\usepackage{graphicx}
\usepackage{booktabs}
\usepackage{color}

\def \hide#1{}

\usepackage{url}
\usepackage[colorlinks]{hyperref}
\hypersetup{%
    plainpages=true,
    breaklinks=true,
    hypertexnames=false,
    pageanchor=true,
    colorlinks=true,
    linkcolor={blue},
    citecolor={red},
    urlcolor={blue},
    anchorcolor={black}
}
\begin{document}
\title{Unconventional Cavity Optomechanics: \\ Nonlinear Control of Phonons in the Acoustic Quantum Vacuum}

\author{Xin Wang}
\affiliation{Institute of Quantum Optics and Quantum Information,
    School of Science, Xi'an Jiaotong University, Xi'an 710049, China}
\affiliation{Theoretical Quantum Physics Laboratory, RIKEN Cluster
    for Pioneering Research, Wako-shi, Saitama 351-0198, Japan}

\author{Wei Qin}
\affiliation{Theoretical Quantum Physics Laboratory, RIKEN Cluster
    for Pioneering Research, Wako-shi, Saitama 351-0198, Japan}

\author{Adam Miranowicz}
\affiliation{Theoretical Quantum Physics Laboratory, RIKEN Cluster
    for Pioneering Research, Wako-shi, Saitama 351-0198, Japan}
\affiliation{Faculty of Physics, Adam Mickiewicz University,
    61-614 Pozna\'n, Poland}

\author{Salvatore Savasta}
\affiliation{Theoretical Quantum Physics Laboratory, RIKEN Cluster
    for Pioneering Research, Wako-shi, Saitama 351-0198, Japan}
\affiliation{Dipartimento di Scienze Matematiche e Informatiche,
    Scienze Fisiche e Scienze della Terra, \\ Universit\`{a} di
    Messina, I-98166 Messina, Italy}

\author{Franco Nori}
\affiliation{Theoretical Quantum Physics Laboratory, RIKEN Cluster
    for Pioneering Research, Wako-shi, Saitama 351-0198, Japan}
\affiliation{Physics Department, The University of Michigan, Ann
    Arbor, Michigan 48109-1040, USA}
\date{\today}

\begin{abstract}
We study unconventional cavity optomechanics and the acoustic
analogue of radiation pressure to show the possibility of
nonlinear coherent control of phonons in the acoustic quantum
vacuum. Specifically, we study systems where a quantized optical
field effectively modifies the frequency of an acoustic resonator.
We present a general method to enhance such a nonlinear
interaction by employing an intermediate qubit. Compared with
conventional optomechanical systems, the roles of mechanical and
optical resonators are interchanged, and the boundary condition of
the phonon resonator can be modulated with an ultrahigh optical
frequency. These differences allow to test some quantum effects
with parameters which are far beyond the reach of conventional
cavity optomechanics. Based on this novel interaction form, we
show that various nonclassical quantum effects can be realized.
Examples include an effective method for modulating the resonance
frequency of a phonon resonator (e.g., a surface-acoustic-wave
resonator), demonstrating mechanical parametric amplification, and
the dynamical Casimir effect of phonons originating from the
acoustic quantum vacuum. Our results demonstrate that
unconventional optomechanics offers a versatile hybrid platform
for quantum engineering of nonclassical phonon states in quantum
acoustodynamics.
\end{abstract}

\maketitle

\affiliation{Institute of Quantum Optics and Quantum Information,
School of Science, Xi'an Jiaotong University, Xi'an 710049, China}
\affiliation{CEMS, RIKEN, Wako-shi, Saitama 351-0198, Japan}

\affiliation{CEMS, RIKEN, Wako-shi, Saitama 351-0198, Japan}
\affiliation{Faculty of Physics, Adam Mickiewicz University,
61-614 Pozna\'n, Poland}

\affiliation{CEMS, RIKEN, Wako-shi, Saitama 351-0198, Japan}
\affiliation{Physics Department, The University of Michigan, Ann
Arbor, Michigan 48109-1040, USA}

\section{Introduction}
\subsection{Conventional optomechanics}
Optomechanical systems~\cite{bowen2015quantum,
aspelmeyer2014cavity}, in which quantized optical fields interact
with a massive movable mirror via radiation pressure, bring
together quantum physics and the macroscopic classical world. The
basic mechanism of optomechanics is that the position of a movable
mirror produces a time-dependent boundary condition of quantized
electromagnetic fields, which in turn modulate an effective cavity
resonant frequency~\cite{Law95}. Optomechanical systems provide a
versatile platform to examine fundamental concepts of quantum
physics and explore the classical-quantum boundary. Examples
include testing wave-function-collapse
models~\cite{Bose97,Bassi13,Blencowe13}, studying the dynamical
Casimir effect
(DCE)~\cite{Schaller02,Kim06,Johansson09L,Wilson2011,Nation12,Johansson13,Macr2018,Stefano18a},
and putting massive objects into nonclassical
states~\cite{Marshall03,Liao16}. One may wonder whether there are
unconventional optomechanical (UOM) systems, where the
\emph{boundary condition of a mechanical resonator can be changed
by a quantized optical field}. We show that such UOM systems can
indeed exist, and give a general method to produce an UOM
nonlinear interaction for controlling phonons (i.e., quantum
engineering of phonons). In particular, we show how to realize: a
mechanical phase-sensitive amplifier,  and  an acoustic analogue
of optical DCE. The acoustic DCE tries to simulate cosmological
phenomena such as Hawking radiation and the Unruh
effect~\cite{Carusotto2009,Boiron15,Eckel18,Schmit2018}. However,
no experiment has shown the phonon DCE using a phonon resonator at
the quantum level: i.e., the effects where quantized photons are
converted into DCE pairs of itinerant phonons.

Let us recall the interaction form in a conventional
optomechanical (COM) system. Setting $\hbar=1$, an exact form of
the interaction Hamiltonian between a single-mode optical field
and a moving mirror is~\cite{Law95}
\begin{equation}
H_{\text{COM}}=G_{\text{COM}}(a+a^{\dagger})^{2}(b+b^{\dagger}),
\label{Hcom1}
\end{equation}
where $a$ and $b$ ($a^{\dagger}$ and $b^{\dagger}$) are the
annihilation (creation) operators of the optical and mechanical
modes, respectively; and the single-photon coupling strength is
$G_{\text{COM}}$. Note that in most studies, the quadratic terms
$a^{2}$ and $a^{\dagger2}$ are often neglected, because they
describe rapidly oscillating virtual processes where photons are
annihilated and created in pairs. As a result, the COM
interaction
\begin{equation}
H'_{\text{COM}}=2G_{\text{COM}}a^{\dagger}a(b+b^{\dagger})
\label{Hcom2}
\end{equation}
is obtained.

\subsection{Main idea of proposed unconventional optomechanics}

\begin{figure}[tbph]
    \centering \includegraphics[width=8.7cm]{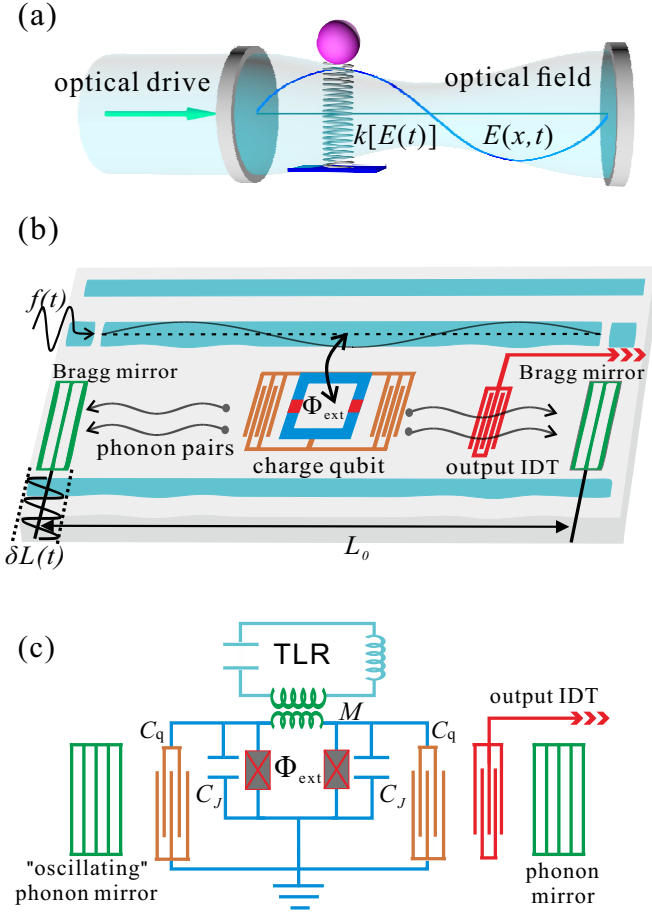}
\caption{(a) Diagrammatic sketch of an unconventional
optomechanical (UOM) system: A localized mechanical oscillator
(with a massive particle) is placed inside an optical cavity, and
its spring constant is linearly modulated by the cavity field
$E(x,t)$. (b) Schematic diagram and (c) lumped-circuit layout
describing an unconventional optomechanical (UOM) system based on a
surface-acoustic-wave (SAW) resonator:: A charge qubit, with two
\allowbreak Josephson junctions (red bars) is placed into the SAW
resonator (confined by two Bragg mirrors). Their interaction is
mediated via two identical inter-digitated-transducers (IDTs) of
capacitance $C_{q}$ via the piezoelectric effects. A
transmission-line resonator (TLR) longitudinally couples to the
charge qubit via the mutual inductance $M$. The UOM Hamiltonian in
Eq.~(\ref{Hint}) can be mapped as the field current operator
$\hat{I}(t)$ effectively changes the SAW resonator boundary
condition. One Bragg mirror acts as a \emph{fast ``oscillating"
mirror.}}
\label{fig1m}
\end{figure}

We start our discussion of an UOM system by considering a
classical mechanical parametric amplifier (MPA), where the spring
constant $k[E(t)]$ of a mechanical oscillator is modulated with a
time-dependent field
$E(t)$~\cite{Rugar91,Carr2000,Zalalutdinov2001,Karabalin2010}. As
depicted in Fig.~\ref{fig1m}(a), we consider that the
amplification source is not a classical drive, but a quantized
electromagnetic field, $E(t)=\varepsilon_{0}(a+a^{\dagger})$,
oscillating at frequency $\omega_{c}$, where $\varepsilon_{0}$ is
the zero-point fluctuation. Expanding the potential term of the
mechanical mode to first order in $E(t)$, we obtain the
interaction term as~\cite{Rugar91}:
\begin{equation}
V(t)=\frac{1}{2}k[E(t)]x^{2}\simeq\frac{1}{2}k_{0}x^{2}+\frac{1}{2}RE(t)x^{2},
\label{vt}
\end{equation}
where $k_{0}$ is the spring constant without the modulating field, and $x=x_{0}(b+b^{\dagger})$ is the mechanical position
operator with $x_{0}$ being the zero-point fluctuation. The last
term in Eq.~(\ref{vt}), describes the response of the spring
constant to the optical field with sensitivity $$
R=\frac{\partial
	k(E)}{\partial E}
\Big|_{E=0},$$ and can be rewritten as
\begin{equation}
H_{\text{UOM}}=G_{\text{UOM}}(b+b^{\dagger})^{2}(a+a^{\dagger}), \quad G_{\text{UOM}}=\frac{R\varepsilon_{0}x_{0}^{2}}{2},
\label{Hint}
\end{equation}
where $G_{\text{UOM}}$ is the
nonlinear coupling strength. Compared with the COM Hamiltonian,
$H_{\text{UOM}}$ has the inverse form, where the roles of the
mechanical oscillator and optical field are interchanged.

We note that we are describing classical quantities using
quantum-mechanical parameters, such as the creation and
annihilation operators in Eqs.~(\ref{Hcom1}) and (\ref{Hcom2}).
Indeed, any classical system can be described using a quantum
formalism. However, this does not mean that the reverse statement
is true. A classical system would not be able to demonstrate a
distinctly quantum performance in an experiment if the
single-phonon coupling strength is suppressed by various kinds of
decoherence processes. Moreover, ultra-weak quantum coherent
signals should be readout with reliable fidelities. In the
following discussions, we will describe in detail methods which
enable the observation of various quantum signatures of UOM
systems.

The UOM interaction requires that the spring constant linearly
responds to a fast-oscillating quantized electric field. To
observe the coherence effects, this response should be
ultra-sensitive (a large $R$) to enable a strong coupling strength
$G_{\text{UOM}}$. Obtaining such large $R$ is still challenging in
conventional MPA experiments. For example, in
Ref.~\cite{Karabalin2010}, the spring constant is modulated by an
external voltage via the piezoelectric \allowbreak effect, and the
sensitivity $R$ is about $40~\text{kHz}/\text{V}$. We assume that
the external voltage is provided by a microwave transmission-line
resonator (TLR), where the typical zero-point voltage fluctuation
is $\sim0.1$--$1~\mu\text{V}$~\cite{Gu2017}. The single-phonon UOM
coupling strength is $G_{\text{UOM}}\simeq
10^{-2}$--$10^{-3}~\text{Hz}$, which is too weak to produce
observable coherent effects.

One can simulate the UOM interaction in a membrane-in-middle
optomechanical system~\cite{Thompson2008, Bruschi2018} based on
the second-order optomechanical interactions and the
semi-classical treatment of a cavity field. We stress that these
membrane-in-the-middle configurations are still based on a COM
system, which does \emph{not interchange} the roles of phonons and
photons. Thus, Refs.~\cite{Thompson2008, Bruschi2018} described
only \emph{simulations} of UOM-type interactions. In contrast to
these works, our proposal describes \emph{real} UOM with the
\emph{interchanged} roles of phonons and photons compared to COM.
Our proposal is based on the first-order optomechanical
interactions and the fully quantum treatment of the cavity field.
Moreover, concerning the proposals in Refs.~\cite{Thompson2008,
Bruschi2018}, no analogous phonon mirror can be found, and it is
hard to control optical quantum fluctuations to reproduce (or
simulate) a controllable boundary condition. Additionally, to
avoid driving the system into the unstable and strongly squeezed
regimes, the simulated interaction strength should be ultra-weak.
So far, no proposal has discussed how to realize a real UOM with
strong enough strength for quantum-optical engineering.

The paper is organized as follows: In Sec.~II and Appendix~A,
we derive the basic qubit-mediated coupling for UOM. In Sec.~III,
we describe UOM based on hybrid superconducting circuits. In
Sec.~IV, we show a few examples of quantum control in UOM. These
include: modulating the frequency of a surface-acoustic-wave
resonator (Sec.~IV~A), quadratic coupling in UOM (Sec.~IV~B and
Appendix~B), and the phonon dynamical Casimir effect (Sec.~IV~C). We
conclude in Sec.~V.

\section{Qubit-mediated coupling for unconventional optomechanics}

Both optical and mechanical oscillators are linear bosonic
systems. Single-phonon-photon nonlinear interactions (e.g., the
COM coupling) are usually much weaker than conventional
light-matter interactions~\cite{Aspelmeyer14,FriskKockum2019}. To
increase their nonlinear interactions, one possible method is to
introduce nonlinear elements. For example, a
Josephson-junction-based qubit can help to induce a strong COM
interaction~\cite{Heik14,Pirkkalainen2015}.

Concerning nonlinearity, a qubit is naturally a highly nonlinear
system. By exploiting this nonlinearity, one may enhance  the UOM
interaction to an observable level. To introduce our idea, we
consider a mechanical oscillator \emph{transversely} interacting
with a qubit with strength $g_{x}$. The optical cavity is involved
in this hybrid system by its \emph{longitudinal} coupling to the
qubit with strength $g_{z}$~\cite{Zhao15,Richer16,Richer17}. By
defining the qubit Pauli operators $\sigma_{z}=|e\rangle\langle
e|-|g\rangle\langle g|$ and $\sigma_{x}=|e\rangle\langle
g|+|g\rangle\langle e|$, where $|g\rangle$ and $|e\rangle$ are the
qubit ground and excited states, the total Hamiltonian includes
three parts, i.e.,
\begin{subequations}
\begin{gather}
H_{T}=H_{0}+H_{1}+H_{2}. \label{Htt}, \\
H_{0}=\frac{1}{2}\omega_{q}\sigma_{z}+\omega_{m}b^{\dag}b+\omega_{c}a^{\dag}a, \\
H_{1}=g_{z}\sigma_{z}\left(a+a^{\dag}\right), \\
H_{2}=g_{x}\sigma_{x}(b+b^{\dagger}).
\end{gather}
\end{subequations}
Here $H_{0}$ is the free Hamiltonian, with $\omega_{q}$, $\omega_{m}$, and
$\omega_{c}$ being the qubit, mechanical resonator mode and
optical cavity mode eigenfrequencies, respectively; while
$H_{1}$ and
$H_{2}$ are the photon-qubit and
phonon-qubit coupling Hamiltonians, respectively. The UOM
interaction can be understood as follows: $H_{1}$ describes the
quantized optical field operator, $\xi=a+a^{\dagger}$, modulating
the qubit frequency as $$\omega_{q}(\xi)\longrightarrow
\omega_{q}+2g_{z} \xi.$$ Moreover, $H_{2}$ leads to a dispersive
coupling between the qubit and the mechanical mode, i.e.,
\begin{equation}
H_{\text{dis}}(\xi)=\chi(\xi)(b^{\dagger}+b)^{2}\sigma_{z}, \quad \chi(\xi)\simeq \frac{g_{x}^{2}}{\omega_{q}(\xi)},
\end{equation}
where the dispersive strength $\chi(\xi)$ is not a constant but depends on the
field operator $\xi$. Different from the standard dispersive form,
$H_{\text{dis}}$ includes the additional quadratic terms
$b^{\dagger2}$ and $b^{2}$, which result from the counter-rotating
terms in $H_{2}$~\cite{Zueco09}. Assuming that $\omega_{q}\gg
\omega_{m}$, we approximately expand  $H_{\text{dis}}(\xi)$ in
$\xi$ to obtain:
\begin{subequations}
    \begin{gather}
    H_{\text{dis}}(\xi)=\sum_{n=0}^{\infty}G_{n} \xi^{n} \sigma_{z}(b^{\dagger}+b)^{2}, \label{Hchix} \\
    G_{n}=\frac{1}{n!}\frac{\partial ^{n}\chi(\xi)}{\partial \xi^{n}}\Big|_{\xi=0}\simeq (-1)^{n}\frac{g_{x}^{2}}{\omega_{q}^{n+1}}(2g_{z})^{n}.
    \label{chiz}
    \end{gather}
    \label{chiUOM}
\end{subequations}
Our detailed derivations are given in Appendix~A. We assume that
the qubit is initially in its ground state $|g\rangle$. Since the
qubit is largely detuned from the mechanical mode, phonons cannot
effectively excite the qubit. Moreover, the longitudinal coupling
commutes with the qubit operator $\sigma_{z}$, which does not
cause any qubit state transition either. Therefore, it is
reasonable to assume that the qubit is approximately in its ground
state with $\langle \sigma_{z}\rangle \simeq-1$ in
Eq.~(\ref{Hchix}). Under these conditions, the zeroth-order term
can be reduced to
\begin{equation}
H_{\text{dis},0}=G_{0}\sigma_{z}(b^{\dagger}+b)^{2}\simeq-2G_{0}
b^{\dagger}b, \label{zerochi}
\end{equation}
which shows that the mechanical frequency is renormalized as
$$\omega_{m}\longrightarrow(\omega_{m}-2G_{0}).$$
The first-
and second-order terms in Eq.~(\ref{Hchix}) describe the
interaction between the mechanical oscillator and the optical
field. The corresponding coupling ratio,
\begin{equation}
\left|\frac{G_{2}}{G_{1}}\right|=\frac{2g_{z}}{\omega_{q}}\ll 1,
\end{equation}
is a small parameter. Therefore, we just consider only the
first-order term. Since the qubit degree of freedom is effectively
eliminated, the Hamiltonian for the mechanical and optical modes
can be written as
\begin{equation}
H_{s}\simeq\omega_{c}a^{\dagger}a+\omega_{m}b^{\dagger}b-G_{1}(b^{\dagger}+b)^{2}
(a^{\dagger}+a). \label{Hs}
\end{equation}

Since the mechanical boundary condition is modulated with a
fast-oscillating optical field, the quadratic terms $b^{2}$ and
$b^{\dagger2}$ in Eq.~(\ref{Hint}) cannot be dropped. To obtain an
exact analog of $H_{\text{COM}}'$, one can employ a low-frequency
optical resonator. Consequently, the rapidly oscillating terms,
describing two-phonon processes, can be neglected. The UOM
interaction can be reduced to
\begin{equation}
H_{\text{UOM}}'=2G_{1}b^{\dagger}b(a+a^{\dagger}). \label{HUOMS}
\end{equation}
As discussed in Appendix~A, an alternative method is to shift the
effective optical frequency via a parametric modulation of the
longitudinal coupling~\cite{Didier15,Cirio17}. Specifically,
assuming that $g_{z}$ is modulated at a frequency close to
$\omega_{c}$, the effective optical frequency
$\omega_{c}'=(\omega_{c}-\omega_{d})$ is shifted, becoming much
smaller than $\omega_{m}$. By controlling the modulating rate
$\omega_{d}$, the optical frequency $\omega_{c}$ is not fixed but
tunable, which makes the UOM system more flexible. Therefore, the
quadratic terms can also be safely neglected, thus, obtaining
Eq.~(\ref{HUOMS}). Similar to the mechanism how radiation pressure
acts on a macroscopic mirror~\cite{Aspelmeyer14},
$H_{\text{UOM}}'$ describes how \emph{acoustic intensity} (or
``phonon pressure'', proportional to $\langle b^{\dagger}b
\rangle$) \emph{induces motions of the optical ``position"
operator} $(a+a^{\dagger})$.
\begin{figure}[tbph]
    \centering \includegraphics[width=8.7cm]{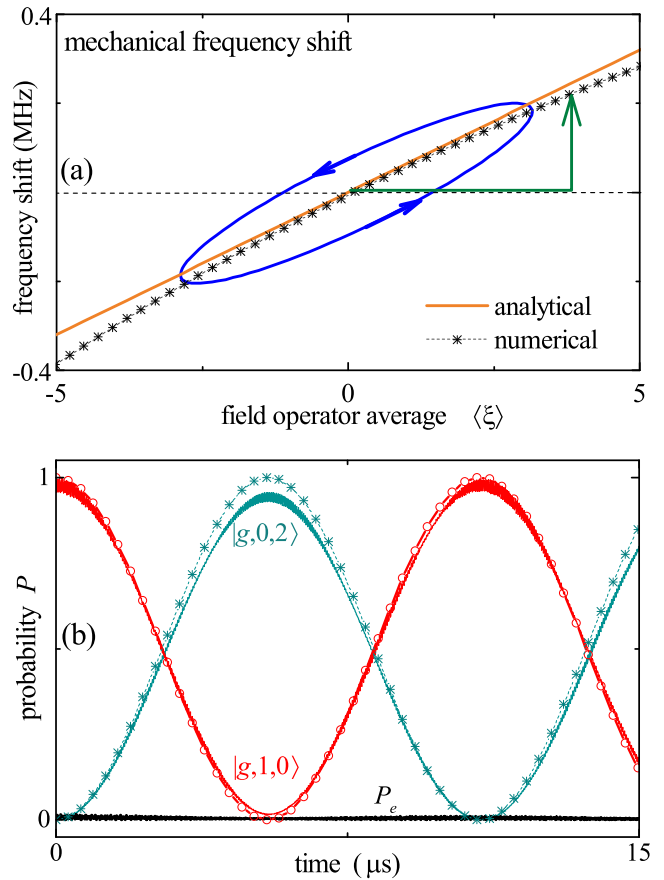}

\caption{(a) Mechanical frequency shift
$(\overline{\omega}_{m}-\omega_{m})$ versus the field operator
average $\langle \xi \rangle=\langle a^{\dagger}+a \rangle$. The
solid (asterisk) curve corresponds to the analytical (numerical)
results. In numerical calculations, the shifted mechanical
frequency is defined as the difference between the first and
second eigenvalues in the subspace of the qubit ground state
$|g\rangle$. The mechanical frequency can be suddenly
(periodically) modulated via with a step longitudinal bias (an
oscillating electromagnetic field) along the green vertical arrow
(the blue loop). (b) The Rabi oscillations between the states
$|g,0,2\rangle$ and $|g,1,0\rangle$. The solid curves (symbols)
represent the evolution described by the original Hamiltonian
$H_{T}$ (the reduced Hamiltonian $H_{\text{Q}}$). $P_{e}$ is the
probability of finding the qubit in its excited state.}
    \label{fig2m}
\end{figure}

\section{Unconventional optomechanics based on hybrid superconducting circuits}

In the previous section, we showed how to obtain the UOM
interaction by employing an intermediate qubit. Although such a
method is very general and not specified to any certain quantum
platform, we now give an example of UOM systems with a surface
acoustic wave (SAW) resonator.

The majority of previous studies about various quantum
features of phonons were demonstrated in localized mechanical
resonators (MRs)(see, e.g., Refs.~\cite{Rabl2009, Poot2010, Teufel11, Viennot2018}),
such as suspended cantilevers and doubly clamped MRs. In such
systems, the size of the MRs is usually of the same order of a phonon
wavelength. Therefore, phonons spread through the whole volume of the MRs
and, thus, it is not easy to observe phonon propagation in those
nanoscale resonators. However, in the studies of SAWs, the
piezoelectric surface behaves as an acoustic waveguide, and is
usually much longer than the phonon wavelength~\cite{Manenti16,
Manenti2017}. As discussed in Refs.~\cite{Gustafsson2014,
Delsing2019}, the SAW propagation effects can be clearly observed
and phonons can interact with atoms placed on their propagation
paths. Due to these features, a SAW resonator can be employed as a
quantum channel to mediate two artificial atoms for quantum
information processing~\cite{Gustafsson2014,Kockum2018,
Delsing2019}.

As shown in Fig.~\ref{fig1m}(b), phonons can be itinerant in a SAW
resonator confined by two Bragg phonon
mirrors~\cite{Schuetz15,Manenti16,Manenti2017,Bolgar18,Kockum2018}.
Similar to an optical cavity, the $N$th resonance acoustic mode
depends on the phonon-mirror distance $L_{0}$ via the relation,
$\omega_{m}/(2\pi)=N v_{e}/(2L_{0})$, where $v_{e}$ is the sound
speed along the crystal surface. Here we consider a charge
qubit~\cite{Gu2017} with two symmetric Josephson junctions (with
Josephson energy $E_{J}$) placed inside the phonon resonator. The
two junctions of the charge qubit form a superconducting quantum
interference device (SQUID), and the total Josephson energy can be
controlled via the external flux $\Phi_{\text{ext}}$ through it.
The qubit capacitance $C_{q}$ of an interdigitated-transducer
(IDT) type shares the same periodicity with the resonator acoustic
mode. The Hamiltonian for the charge qubit can be expressed
as~\cite{You2003,Gu2017}:
\begin{equation}
H_{\text{q}}=4E_{C}(\hat{n}-n_{g})^{2}-2E_{J}\cos\!\left(\frac{\pi\Phi_{\text{ext}}}{\Phi_{0}}\right)\cos{\phi},
\end{equation}
where $E_{C}=e^{2}/(2C_{\Sigma})$ is the total charging energy of
the two junctions, and $C_{\Sigma}=C_{J}+C_{g}$ with $C_{J}$ being
the Josephson capacitance. To suppress the charge noise, one can
apply a dc voltage to bias at the charge degeneracy point
$n_{g}=1/2$. The charge qubit Hamiltonian takes the
form~\cite{Irish03,Gu2017}:
\begin{eqnarray}
H_{\text{q}}&\simeq&-4E_{C}\delta n_{g} (|1\rangle\langle 1|-|0\rangle\langle 0|) \notag \\
&&-E_{J}\cos\!\left(\frac{\pi\Phi_{\text{ext}}}{\Phi_{0}}\right)(|1\rangle\langle
0|+|0\rangle\langle 1|), \label{eqN}
\end{eqnarray}
where $|0\rangle$ and $|1\rangle$ are the charge qubit states, and
$\delta n_{g}=C_{q}V(t)/(2e)$ is the offset charge deviation from
the optimal point, with $V(t)$ being the external voltage drive.
The qubit capacitance $C_{q}$ serves as a coupling element between
the SAW resonator and the charge qubit. The interaction mechanism
can be understood as follows: An acoustic wave travels on the
crystal surface, and generates an oscillating voltage $V(t)$ on
the qubit capacitance $C_{q}$ due to the piezoelectric
effect~\cite{Manenti2017}. Note that $V(t)$ is induced by the
quantized motion and can be viewed as a time-dependent drive on
the charge qubit. Although many discrete acoustic modes can, in
principle, be excited in the phonon resonator, only one central
mode is strongly coupled to the qubit~\cite{Manenti2017}. Thus, it
is physically justified to consider a single acoustic mode here.
The voltage difference is associated with the zero-point
mechanical fluctuation $u_{0}$ as $V(t)=u_{0}(b+b^{\dagger})$. In
the rotated basis with $$|e\rangle=\frac{|1\rangle-|0\rangle}{\sqrt{2}}, \quad  |g\rangle=\frac{1\rangle+|0\rangle}{\sqrt{2}},$$ the SAW-qubit
coupling can be approximately written as~\cite{Manenti2017}:
\begin{equation}
H_{\text{qm}}=-4E_{C}\delta
n_{g}\sigma_{x}=-\frac{eC_{q}}{C_{\Sigma}}u_{0}(b+b^{\dagger})\sigma_{x},
\label{Hqmmain}
\end{equation}
which corresponds to the transverse coupling between the SAW
resonator and the  charge qubit.

As depicted in Fig.~\ref{fig1m}(c), the effective Josephson energy
depends on the external flux bias $\Phi_{\text{ext}}$. One can
couple the charge qubit with a TLR via a mutual inductance $M$.
The central conductor of the TLR is along the $x$ direction, and
the interaction position is assumed to be at an anti-node point of
the current field $\hat{I}(x,t)$. The current
$\hat{I}=I_{0}(a+a^{\dagger})$ creates a flux perturbation $$\delta
\Phi_{\text{ext}}=MI_{0}(a+a^{\dagger})$$ on the qubit static bias
flux $\Phi_{\text{ext}}^{0}$, where $I_{0}$ is the current
zero-point-fluctuation amplitude, and $a$ ($a^{\dagger}$) is the
annihilation (creation) operator of the microwave photons.
Therefore, we can expand the Josephson term in Eq.~(\ref{eqN}), to
first order in $\delta \Phi_{\text{ext}}$, and
obtain~\cite{Sun06}:
\begin{gather}
H_{q}=\frac{\omega_{q}}{2}\sigma_{z}+ \frac{1}{2}\!\left( \frac{\partial \omega_{q}}{\partial \Phi_{\text{ext}}} \right)\Big|_{\Phi_{\text{ext}}^{0}} \sigma_{z}\delta \Phi_{\text{ext}},   \label{Hqex} \\
\omega_{q}=2E_{J}\cos\!\left(\frac{\pi\Phi_{\text{ext}}}{\Phi_{0}}\right).
\end{gather}
Note that the second term in Eq.~(\ref{Hqex}) describes the
interaction between the charge qubit and the TLR, and can be
written as
\begin{equation}
H_{\text{qc}}=g_{z}(a+a^{\dagger})\sigma_{z},
\end{equation}
where the longitudinal coupling strength is
\begin{equation}
g_{z}=-\frac{\pi
E_{J}}{\Phi_{0}}\sin\!\left(\frac{\pi\Phi_{\text{ext}}}{\Phi_{0}}\right)MI_{0}.
\end{equation}

Up to now, we considered $g_{z}$ to be a constant. A
\emph{parametrically modulated longitudinal coupling} between a
superconducting qubit and a $\lambda/4$ (quarter wavelength) TLR
can be realized by applying a time-dependent flux through the
SQUID loop. A detailed description of the method can be found in
Ref.~\cite{Didier15}.

In Refs.~\cite{Manenti2017,Bolgar18}, the SAW resonator is assumed
to couple with a transmon, and their coupling shares a similar
expression as in Eq.~(\ref{Hqmmain}) for the SAW-qubit coupling,
except for a dimensionless parameter. The UOM interaction mediated
by a transmon can also be produced. However, different from the
charge qubit, a transmon is a weakly anharmonic system, and the
UOM interaction based on a dispersive coupling (as described in
Sec. II) will be disturbed by an imperfect state truncation at the
first-excited level~\cite{Sete2015,Burillo16}. The UOM interaction
in Eq.~(\ref{Hchix}) should be derived by considering
higher-energy levels.

In addition to the SAW resonator, phonons can also exist in a
localized MR. As discussed in
Ref.~\cite{Viennot2018}, the transverse-coupling strength between a
drum-type mechanical oscillator and a charge qubit can be
engineered in the strong-coupling regime (about tens of MHz). By
coupling the TLR (or \textit{LC}-resonator) quantized
electromagnetic field with the split-junction loop of the charge
qubit, the required longitudinal interaction can also be
induced~\cite{Sun06}. Therefore, one can realize the required
Hamiltonian Eq.~(\ref{Htt}), to generate the qubit-mediated UOM
interaction [Eq.~(\ref{Hs})] for both SAW resonators and localized
MRs.

\section{Quantum control in unconventional optomechanics}

The proposed UOM system differs from a COM system as follows:
First, \emph{the phonon-resonator boundary condition is modulated
via a fast-oscillating optical field.} Therefore, we cannot drop
the quadratic term (as we do for a COM system), which can induce
observable quantum effects. Second, the COM Hamiltonian results
from radiation pressure. However, the UOM interaction is mediated
via a qubit, and has no relation to moving the massive phonon
mirrors via an optical (microwave) field. Indeed, it originates
from a dispersive coupling being modulated by a quantized optical
field. Due to these differences, the UOM interaction enables to
observe some unconventional quantum phenomena. Below, by
considering the SAW-resonator-based UOM system as an example, we
show some possible quantum-control applications.

\subsection{Modulating the frequency of a surface-acoustic-wave resonator}

Using a large mutual inductance, the TLR-qubit interaction
strength can easily enter into the strong- or even
ultra-strong-coupling regimes~\cite{Gu2017,FriskKockum2019}. The
SAW-qubit coupling reaches tens of $\text{MHz}$ in
experiments~\cite{Manenti2017,Bolgar18}. By setting
$g_{x}/(2\pi)=60~\text{MHz}$, $g_{z}/(2\pi)=40~\text{MHz}$,
$\omega_{q}/(2\pi)=3~\text{GHz}$, and
$\omega_{m}/(2\pi)=250~\text{MHz}$, one can obtain a single-phonon
UOM coupling strength $G_{1}/(2\pi)\!\simeq-32~\text{kHz}$, which
is significantly enhanced compared to a direct coupling via the
piezoelectric effect. We assume these values in our numerical
calculations~\cite{Johansson13qutip,Johansson12qutip}.

Analogously to COM, the optical field operator modifies the
mechanical frequency as
\begin{equation}
\overline{\omega}_{m}=\sqrt{\omega_{m}^{2}-4G_{1}\omega_{m}\langle
    a^{\dagger}+a\rangle}.
\end{equation}
To show this, we diagonalize
Eq.~(\ref{Htt}), and plot $(\overline{\omega}_{m}-\omega_{m})$
versus $\langle \xi \rangle=\langle a^{\dagger}+a \rangle$ in
Fig.~\ref{fig2m}(a). Clearly, the effective mechanical frequency
is shifted away from $\omega_{m}$ by increasing $\langle \xi
\rangle$. To derive $H_{\text{UOM}}$ in Eq.~(\ref{Hint}), we just
expand to first order in $\langle \xi \rangle$, by assuming
$2g_{z}\langle \xi \rangle \ll \omega_{q}$. Therefore, our
analytical formula for the shifted frequency slightly differs from
the numerical results when $\langle \xi \rangle\gg 1$.

Note that the TLR can be replaced by a 1D microwave
guide~\cite{Gu2017}, to which a classical microwave field can be
applied. We assume that the current drive signal is $$I(t)=\Theta
(t)I_{c},$$ where $\Theta (t)$ is the Heaviside unit step function,
and $I_{c}$ is the dc current strength applied to the 1D microwave
guide when $t\geq0$. The longitudinal coupling should be replaced
with the classical step drive $$H_{\Theta}=\Omega_{s}
\sigma_{z}\Theta(t),$$ where
\begin{equation}
\Omega_{s}=-\frac{\pi
E_{J}}{\Phi_{0}}\sin\!\!\left(\frac{\pi\Phi_{\text{ext}}}{\Phi_{0}}\right)M
I_{c}. \label{current_o}
\end{equation}
Following the derivations in Sec.~I and the discussions in
Ref.~\cite{Law95}, the interaction form corresponds to
\emph{changing the frequency and the length of the SAW resonator}
at $t=0$ with amounts
\begin{equation}
\delta \omega_{m}=\frac{4g_{x}^{2}}{\omega_{q}^{2}}\Omega_{s},
\quad \delta L=\frac{\delta \omega_{m}}{\omega_{m}}L_{0},
\label{deltaL}
\end{equation}
respectively. Therefore, to effectively modify the resonance
frequency of the SAW resonator, one can simply apply a current
bias on the qubit longitudinal degree of freedom.

\subsection{Quadratic coupling for mechanical parametric amplification}

The phonon-resonator boundary condition in an UOM system is
modulated by an optical field at an ultrahigh rate, which can
easily exceed the phonon-resonator frequency. By setting
$\omega_{c}=2\omega_{m}$, one can reduce $H_{\text{UOM}}$ to a
simpler Hamiltonian
\begin{equation}
H_{\text{Q}}=G_{1}(a^{\dagger}b^{2}+ab^{\dagger 2}).
\end{equation}
We denote
$|g (e),n,m\rangle$ for the system containing $n$ photons and $m$
phonons, with the qubit in its ground (excited) state. The
quadratic interaction $H_{\text{Q}}$ can be verified clearly from
Fig.~\ref{fig2m}(b), which shows the Rabi oscillations between the
states $|g,0,2\rangle$ and $|g,1,0\rangle$ (solid curves, governed
by $H_{T}$). The curves with symbols correspond to $H_{\text{Q}}$,
which well match the exact results. During this photon-phonon
conversion, the qubit-excited probability $P_{e}$, oscillates with
an ultralow amplitude above zero (the black curve). Thus, we can
safely neglect the qubit degree of freedom when deriving the
effective UOM Hamiltonian. Note that $H_{\text{Q}}$ can mimic the
optomechanical quadratic-interaction in a membrane-in-middle
cavity system~\cite{Thompson2008,Sankey2010,Liao2014} with a much
stronger strength (enhanced by the qubit). Therefore, various
quantum-control applications, such as photon blockade and the
generation of macroscopically distinct superposition state
(Schr\"odinger cat-like
states)~\cite{Nunnenkamp10,Tan13,Liao2013}, can be realized.

Given that the longitudinal drive of the qubit is
\begin{equation}
H_{d}=\Omega_{d} \sigma_{z}\cos(2\omega_{m}t+\phi),
\end{equation}
we can approximately replace the field operator with a classical
amplitude, i.e.,
$$a\rightarrow\alpha=\left(\frac{2g_{x}}{\omega_{q}}\right)^{2}\Omega_{d}\exp(-i\phi).$$
\begin{figure}[tbph]
	\centering \includegraphics[width=8.7cm]{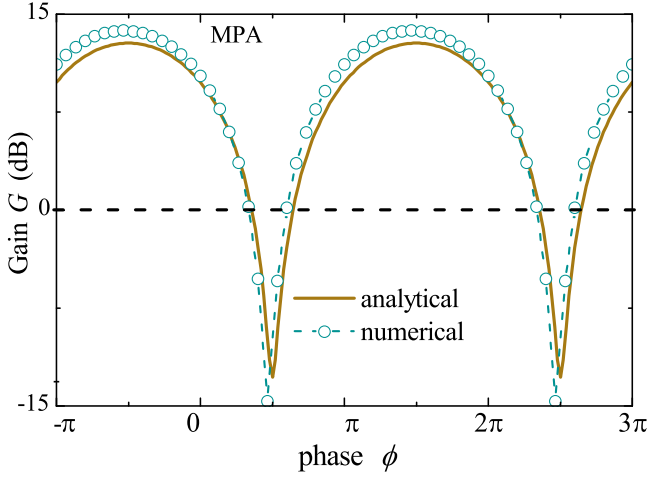}
	\caption{The mechanical gain rate $G$ of the input signals changing with the relative phase $\phi$ for MPA. The analytical results correspond to Eq.~(\ref{gainG}). We set $\alpha/(2\pi)=0.045~\text{MHz}$. The qubit and phonon decay rates are $\Gamma/(2\pi)=0.05~\text{MHz}$ and $\kappa/(2\pi)=0.2~\text{MHz}$, respectively.}
	\label{fig3m}
\end{figure}

When an ultraweak mechanical signal is injected into this UOM
system, one quadrature $X_{\text{in}}$ is amplified to an output
$X_{\text{out}}$ without introducing extra noise~\cite{Caves1982}.
Note that the mechanism of an MPA based on a qubit-mediated UOM
system is novel and different from that in conventional MPAs,
where we should modulate in time the spring constant. In
Appendix~B, we derive the MPA gain as follows~\cite{Clerk10}:
\begin{equation}
G(\phi)=\left|\frac{X_{\text{out}}}{X_{\text{in}}}\right|=\frac{16|\alpha|^{2}+\kappa^{2}-8|\alpha|\kappa
\sin{\phi}}{16|\alpha|^{2}-\kappa^{2}}. \label{gainG}
\end{equation}
In experiments, only some mechanical oscillators can be fabricated
with a tunable spring constant~\cite{Karabalin2010}. Based on our
proposed UOM mechanism, we find another general method for the MPA
process by applying a longitudinal driving on the coupled qubit.

In Fig.~\ref{fig3m}, we plot $G$ versus $\phi$, and find that the
gain reaches its maximum $G\simeq14~\text{dB}$ at $\phi=-\pi/2$.
Different from an ideal MPA, there will be an qubit-induced Kerr
nonlinearity. As discussed in Appendix~B, once plenty of phonons
are injected into the SAW resonator, the Kerr nonlinearity will
destroy the amplifying process and drive the system out of the
quasi-dispersive
regime~\cite{Blais04,Boissonneault08,Boissonneault09}. To avoid
these undesired mechanisms, we require that the maximum phonon
number satisfies Eq.~(\ref{B9}). Therefore, the maximum gain is
bounded by:
\begin{equation} G_{c}\simeq(2N_{c}+1)\left(1+\sqrt{\frac{2N_{c}}{2N_{c}+1}}\right)^{2}\simeq(8N_{c}+4),
\label{gainmax}
\end{equation}
Once the phonon number is beyond $N_{c}$, the qubit can be excited
effectively, and the qubit-mediated UOM model is not valid any
more. However, even in our case with strong coupling strengths,
the critical gain is still $G_{c}\simeq29~\text{dB}$, below which
one can safely amplify the mechanical signal for quantum
measurements ~\cite{Nunnenkamp10,Clerk10}.

\subsection{Phonon dynamical Casimir effect}

Since the SAW frequency can be modulated by an optical field along
the blue loop as shown  Fig.~\ref{fig2m}(a)~\cite{Law95}, one can
alter the mode intensity of the \emph{acoustic quantum vacuum} by
the optical field~\cite{Nation12}. Analogously to the optical DCE
with an optomechanical system, in the following we describe the
\emph{phonon} DCE based on the UOM
system~\cite{Jaskula12,Motazedifard2017}.

As shown in Fig.~\ref{fig1m}(b), the interaction given by
Eq.~(\ref{Hs}) can be interpreted as photons changing the
effective distance $L_{0}$ between two phonon
mirrors~\cite{Law95}. This indicates that the electromagnetic
field can alter the mode intensity of the \emph{phonon} field.
Analogously to the photon dynamical Casimir effect
(DCE)~\cite{Johansson09L,Johansson09,Wilson2011}, phonon pairs are
emitted due to the modulated-boundary condition of the
\emph{acoustic quantum vacuum.}

To verify this, we assume that the intracavity phonons can escape
from the SAW resonator from an output channel. As shown in
Fig.~\ref{fig1m}(c), one can employ an IDT to convert output phonons into electromagnetic signals
(phonons)~\cite{Manenti2017}. The boundary condition for the
output field $c(t)$ and the intracavity SAW field $b(t)$ is\
\begin{equation}
 c(t)=b_{\text{in}}(t)+\sqrt{\kappa}b(t),
\end{equation}
where $\kappa$ is the phonon escape rate from the SAW resonator,
and $b_{\text{in}}(t)$ is assumed to be the vacuum input field.
The output-phonon number per second is expressed as
$P_{\text{out}}=\kappa \langle c^{\dagger}c\rangle$. To describe
the correlations between phonons, we define the second-order
correlation function of the output field as
\begin{equation}
g_{2}(\tau)=\lim_{t \to \infty} \frac{\langle c^{\dagger}(t)
c^{\dagger}(t+\tau)c(t+\tau)c(t)  \rangle}{\langle
c^{\dagger}(t)c(t)\rangle^{2}},
\end{equation}
with $\tau$ being the delay time. The phonon-flux spectrum density
(detected by IDTs) is defined as
\begin{equation}
S(\omega)=\text{Re}\int_{0}^{\infty}\langle
c^{\dagger}(0)c(\tau)\rangle e^{i\omega \tau}d \tau
\end{equation}
We note that a phonon power spectrum can also be measured via
electromotive techniques (see, e.g., Ref.~\cite{Liu10} and
reference therein). According to the Wiener-Khinchin
theorem~\cite{Scully1997}, and by replacing the output operator
with the intracavity field, one can find that
\begin{equation}
S(\omega)\propto n_{\text{out}}(\omega)=\int_{0}^{\infty}
\text{Tr}[\rho b^{\dagger}(\omega)b(\omega')]d\omega',
\end{equation}
where
$$b(\omega)=\frac{1}{\sqrt{2\pi}}\int_{-\infty}^{\infty}b(t)e^{-i\omega
t}dt$$ is the Fourier transform of the intracavity field operator
$b(t)$, and satisfies the canonical commutation relation
$$[b(\omega), b^{\dagger}(\omega')]=\delta(\omega-\omega').$$

Assuming that the TLR is resonantly driven via a coherent field
with strength $\epsilon$, i.e.,
\begin{equation}
H_{d}(t)=\epsilon[a
\exp(i\omega_{c}t)+a^{\dagger}\exp(-i\omega_{c}t)].
\end{equation}
We numerically simulate the quantum evolution of the system described
by the Lindblad-type master equation
\begin{eqnarray}
\frac{d\rho (t)}{dt}&=&-i[H_{1}+H_{d}(t),\rho (t)]+\Gamma D[\sigma_{-}]\rho(t)+ \gamma D[a]\rho(t)  \notag  \\
&&+\kappa n_{\rm{th}} D[b]\rho(t)+\kappa(n_{\rm{th}}+1)
D[b^{\dag}]\rho(t), \label{mastereq}
\end{eqnarray}
where $\Gamma$, $\gamma$, and $\kappa$ are the decay rates for the
qubit, optical resonator, and SAW resonator, respectively,
$n_{\rm{th}}$ is the thermal phonon number, and $$D[A]\rho =(2A\rho
A^{\dag }-A^{\dag}A\rho -\rho A^{\dag}A)/2$$ is the decoherence
term in the Lindblad superoperator form. Note that we consider the
original Hamiltonian $H_{T}$, given in Eq.~(\ref{Htt}) (including
the qubit degree of freedom), rather than the reduced UOM
Hamiltonian $H_{s}$, given in Eq.~(\ref{Hs}).

\begin{figure}[tbph]
	\centering \includegraphics[width=8.9cm]{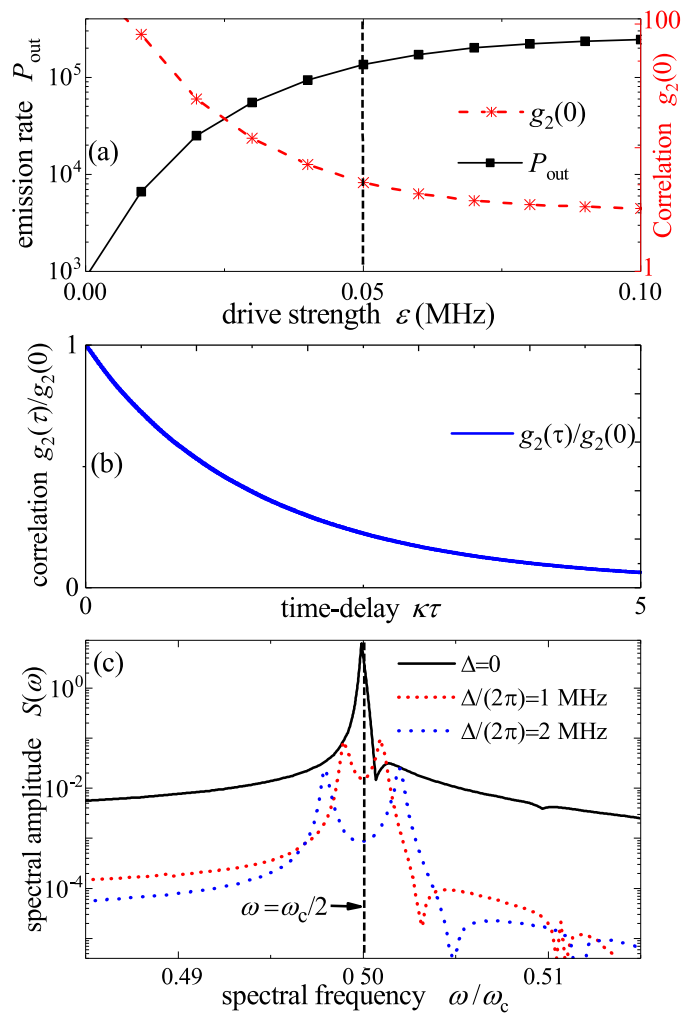}
	\caption{(a) Phonon-output rate $P_{\text{out}}$, and correlation function $g_{2}(0)$, as functions of the TLR drive strength $\epsilon$. The dashed line is plotted at $\epsilon/(2\pi)=0.05~\text{MHz}$, which is set in plots (b) and (c). (b) Normalized correlation function $g_{2}(\tau)/g_{2}(0)$ changing with  the delay time $\kappa \tau$. (c) Phonon output spectrum density $S(\omega)$, for different detuning cases. The dotted line position corresponds to $\omega=\omega_{c}/2$. Here the decay rates are: $\Gamma/(2\pi)=0.05~\text{MHz}$, $\kappa/(2\pi)=0.2~\text{MHz}$, and $\gamma/(2\pi)=0.1~\text{MHz}$.}
	\label{fig4m}
\end{figure}

We first consider the SAW resonator coupled to a zero-temperature
reservoir ($n_{\rm{th}}=0$). As predicted in the optical DCE, with
increasing the drive strength $\epsilon$, the effective length of
the SAW resonator is modulated with a higher amplitude. In
Fig.~\ref{fig4m}(a), by setting $\omega_{c}=2\omega_{m}$, we plot
the phonon output rate $P_{\text{out}}$ changing with $\epsilon$.
We find that, with large $\epsilon$, $P_{\text{out}}$ is enhanced.
At $\epsilon/(2\pi)=0.05~\text{MHz}$ (dashed line), the output
phonon number per second is about $P_{\text{out}}\backsimeq1\times
10^{5}$, which can be effectively detected by an IDT
measurement~\cite{Bolgar18}.

Since there are no thermal excitations and no coherent drive
applied to the SAW resonator, these phonons might be generated by
the phonon DCE, and have the same quantum signatures as photons
generated in the optical DCE. As discussed in
Refs.~\cite{Wilson2011,Lhteenmki2013}, the DCE excitations are
created in pairs. To show this, in Fig.~\ref{fig4m}(a), we plot
the second-order correlation function $g_{2}(0)$ as a function of
the TLR drive strength $\epsilon$. We find that the generated
phonons have super-Poissonian phonon-number statistics with
$g_{2}(0)\gg2$. When $\epsilon$ has ultralow amplitude, $g_{2}(0)$
becomes infinitely large~\cite{Johansson09}. Due to the increasing
phonon intensity~\cite{Macr2018}, $g_{2}(0)$ decreases with
increasing the drive strength. Moreover, in Fig.~\ref{fig4m}(b),
we plot the normalized correlation function $g_{2}(\tau)/g_{2}(0)$
changing with the delay time $\kappa\tau$. We find that
$g_{2}(0)\gg g_{2}(\tau)$, indicating that the emitted phonons
exhibit strong bunching, which is due to the same mechanism as
that in the optical DCE~\cite{Johansson09}.

In Fig.~\ref{fig4m}(c), we plot the spectrum density
$n_{\text{out}}(\omega)$ for different values of the detuning
$\Delta_{d}=\omega_{c}-2\omega_{m}$. When $\Delta_{d}=0$, the
phonon-flux density spectrum shows a single-peak at
$\omega=\omega_{s}/2$. When $\Delta_{d}\neq 0$
[Fig.~\ref{fig4m}(a)], $n_{\text{out}}(\omega)$ shows a clearly
symmetric bimodal spectrum at $\omega'$ and $\omega''$, with
$\omega'+\omega''=\omega_{c}$, which is another strong indication
of the phonon DCE~\cite{Lambrecht96}. Compared with the resonance
case, the correlated emitted phonon pairs are not degenerate any
more, but distributed into two conjugate modes with different
frequencies. When increasing the detuning $\Delta_{d}$, the peaks
of two modes separate with larger distance, while the amplitudes
are suppressed significantly.
\begin{figure}[tbph]
    \centering \includegraphics[width=8.9cm]{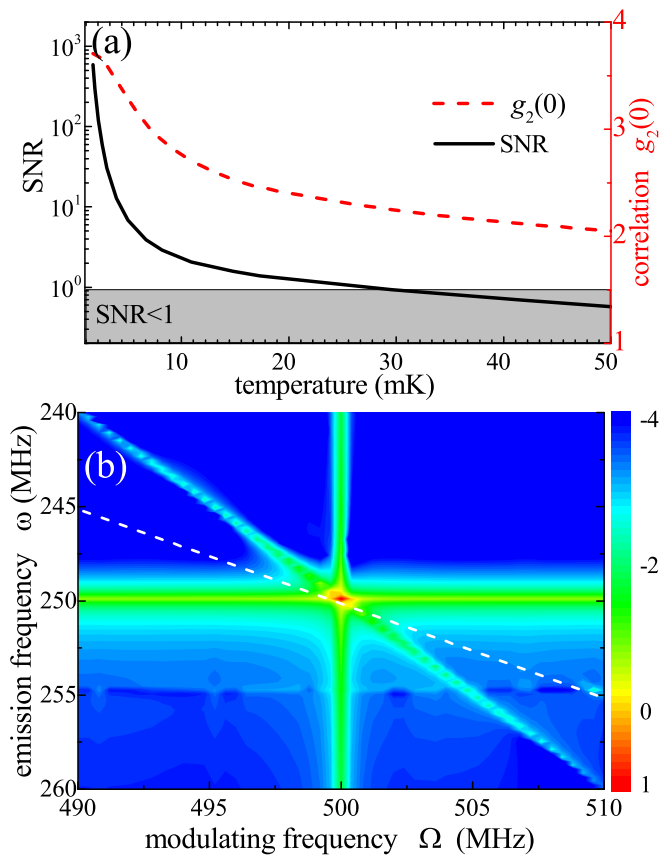}
    \caption{(a) Signal-to-noise-ratio (SNR) and second-order correlation function $g_2(0)$ versus the thermal phonon number $n_{\text{th}}$. The grey area corresponds to $\text{SNR}<1$. (b) The emitted phonon spectrum $\log[S(\omega)]$ changes with the modulating frequency $\Omega$.  A bimodal spectrum structure distributes along the dashed line (which corresponds to $\omega=\Omega/2$). Here we set the coherent drive amplitude on the qubit as $\Omega_{d}/(2\pi)=100~\text{MHz}$. Other parameters are the same as those in Fig.~\ref{fig4m}.}
    \label{fig5m}
\end{figure}

As discussed in Sec. II, the TLR can be replaced by a 1D
transmission waveguide, which can support a classical drive
current. Equations~(\ref{current_o}) and (\ref{deltaL}) indicate
that the pulse shape in the waveguide directly determines how the
effective phonon resonator length changes with time. We consider a
simple case where a sinusoidal current pulse $$I(t)=\Theta
(t)I_{c}\cos(\Omega t)$$ is applied, which is equal to a coherent
drive on the qubit operator $\sigma_{z}$ with strength
$\Omega_{d}=g_{z}I_{c}/I_{0}$. Following Eq.~(\ref{deltaL}),
\emph{the current produces a time-dependent modulation of the
effective phonon resonator length}
\begin{equation}
\delta
L(t)=\frac{4g_{x}^{2}}{\omega_{q}^{2}}\frac{\Omega_{d}}{\omega_{m}}L_{0}\cos(\Omega
t).
\end{equation}
The sinusoidal modulation can be mapped as \emph{moving a phonon
mirror with a non-uniform acceleration}~\cite{Nation12}.
Similarly, we can observe the \emph{phonon DCE}. In the following,
we consider how a non-zero temperature phonon reservoir
influencing the \emph{phonon} DCE signals. Compared with
photons, phonons are much more fragile to environmental noise.
Thus, their generation and detection processes should be
considered carefully. Especially, the DCE signals should be
distinguished from thermal excitations and other kinds of noise.
In the following discussions, we present methods to identify the
DCE phonons from the sea of noisy thermal phonons.

As shown in Fig.~1(b), the phonons in the SAW resonator can
be collected by the output IDT channel. Once considering thermal
effects, the emitted phonons from the SAW resonator can be divided
into two types: the DCE-induced phonon pairs and thermalized
incoherent phonons.

One can define the signal-to-noise-ratio (SNR) of the DCE as
\begin{equation}
\text{SNR}=\frac{P_{\text{out}}-P_{\text{out}}^{\text{th}}}{P_{\text{out}}^{\text{th}}},
\end{equation}
where $P_{\text{out}}^{\text{th}}$ is the output \emph{phonon}
rate without modulating the effective resonator length, and
$P_{\text{out}}$ is the total output phonon rate. Therefore,
$P_{\text{out}}^{\text{th}}$ corresponds to purely thermal
excitations and should be considered as a noise contribution. In
Fig.~5(a), by setting $\Omega=2\omega_{m}$, we plot the SNR of the
DCE versus the environment temperature $T$. We find that, the SNR
decreases quickly with increasing temperature. When
$T>27~\text{mK}$, the $\text{SNR}$ is below 1, indicating that the
DCE signal is hidden in the noisy background.

In addition to employing the output power to confirm the DCE
phonons, we can detect their quantum correlations. As discussed
before, the DCE phonons are generated in pairs. Therefore, their
second-order correlation function $g_{2}(0)$ is super-Poissonian,
and much higher than that of thermal excitations. In
Fig.~\ref{fig5m}(a), we plot $g_{2}(0)$ versus temperature $T$. We
find that $g_{2}(0)$ is very large, with $g_{2}(0)\gg2$ in the
limit of $T\rightarrow 0$. When $T\simeq 50~\text{mK}$,  the DCE
phonons are significantly polluted by thermal excitations, and the
output correlation function $g_{2}(0)\simeq 2$ is almost the same
as that for genuine thermal noise. The numerical results shown in
Fig.~\ref{fig5m}(a) indicate that the DCE signal can be well
separated from thermal noise given that the environment
temperature is below $T\simeq 20~\text{mK}$. The hybrid quantum
circuits based on SAWs are usually placed in diluted
refrigerators, in which the temperature $T\simeq 20~\text{mK}$ is
achievable~\cite{Gu2017, Satzinger2018}. At temperatures of tens
of mK, the quantum manipulating and topography measurement of
propagating SAW phonons have been realized
experimentally~\cite{Gustafsson2014, Bolgar18, Manenti2017,
Satzinger2018}. Therefore, we believe that the observation of
these quantum signatures of DCE phonon pairs is possible using
current experimental approaches.

In Fig.~\ref{fig5m}(b), by setting $n_{\text{th}}=0.02$, we plot
the emission spectrum $S(\omega)$ changing with the coherent
modulation frequency $\Omega$. The dashed line position is plotted
for $\omega=\Omega/2$. The bimodal structure of the emission
spectrum is still kept. However, due to thermal noise, the two
peaks are not symmetric any more. To suppress the effect of
thermal noise on the DCE signals, one can employ high-frequency
modes of the SAW cavity. As discussed in the experiments reported
in Refs.~\cite{Manenti16,Manenti2017,Bolgar18}, the SAW resonance
frequency can be engineered for about several GHz, and the thermal
occupation number $n_{\text{th}}$ can be below $10^{-3}$ at
temperatures $\sim20$ mK, which is within the capability of
up-to-date hybrid quantum circuit implementations in dilution
refrigerators.

In Fig.~\ref{fig2m}(b), the time-dependent evolutions indicates
that there is the vacuum Casimir-Rabi coupling in UOM
systems~\cite{Macr2018}, and we can the observe the phonon DCE in
an UOM system. Figure~\ref{fig4m}(c) is also another strong indication of the
\emph{phonon DCE} where excitations are created in pairs
[Fig.~\ref{fig1m}(b)]. Compared with the resonance case, the
emitted phonon pairs in the detuning cases are not degenerate any more, but distributed
into two conjugate modes with different frequencies.

To observe the \emph{optical DCE}: (i) the mechanical mode should
oscillate at an ultra-high frequency (corresponding to moving the
mirror near the speed of light, instead of the speed of sound for
the phonon case), and (ii) a strong optomechanical interaction
should be induced~\cite{Johansson09L,Wilson2011,Nation12}. Both
requirements are exceedingly challenging in experiments. So far,
no experiment has successfully demonstrated a real optical DCE
involving the mechanical-optical energy
conversion~\cite{Macr2018}. However, the \emph{phonon DCE}
described here is much easier to induce and observe, since the
boundary condition is modulated by a microwave field: The
electromagnetic frequency can easily overwhelm the phonon
resonator frequency. Moreover, a strong mechanical-optical UOM
coupling, which is enhanced by the intermediate qubit, enables
observing the phonon DCE at the quantum level.

\section{Conclusions}

We proposed an UOM mechanism, which describes how the frequency of
a mechanical mode is effectively modulated by a quantized optical
field. We presented a general method to enhance the UOM coupling
via an intermediate qubit. For example, by considering a SAW
resonator, we found that the effective resonator length is not
fixed, but can be shifted in a large range by simply applying a
longitudinal bias on the qubit, which allows more controllability
in SAW-resonator experiments. In principle, analogous of various
quantum effects studied in COM, can be demonstrated in UOM
systems, but with the interchanged roles of \allowbreak photons
and phonons. Recently, quantum acoustodynamics has emerged as a
powerful platform to explore quantum features of acoustic waves.
The UOM mechanism allows to manipulate itinerant phonons at the
quantum
level~\cite{Schuetz15,Manenti16,Manenti2017,Bolgar18,Kockum2014}.
For example, an UOM system can serve as a \emph{nonlinear
transducer converting quantum information between acoustic waves
and microwave resonators}. Other examples include: mechanical
phase-sensitive amplification and phonon DCE. We hope that even
other quantum mechanisms and applications can be developed in UOM
systems in future studies.

\section*{Acknowledgments}

The authors acknowledge fruitful discussions with Drs.~Anton Frisk
Kockum and Sergey Shevchenko. X.W. is supported by the China
Postdoctoral Science Foundation No.~2018M631136, and the Natural
Science Foundation of China (Grant No.~11804270). F.N. is supported in part by the:
MURI Center for Dynamic Magneto-Optics via the
Air Force Office of Scientific Research (AFOSR) (FA9550-14-1-0040),
Army Research Office (ARO) (Grant No. Grant No. W911NF-18-1-0358),
Asian Office of Aerospace Research and Development (AOARD) (Grant No. FA2386-18-1-4045),
Japan Science and Technology Agency (JST) (via the Q-LEAP program, and the CREST Grant No. JPMJCR1676),
Japan Society for the Promotion of Science (JSPS) (JSPS-RFBR Grant No. 17-52-50023, and
JSPS-FWO Grant No. VS.059.18N), the RIKEN-AIST Challenge Research Fund,
the Foundational Questions Institute (FQXi), and the NTT-PHI Lab.

\section*{APPENDICES}
\setcounter{equation}{0}
\renewcommand{\theequation}{A\arabic{equation}}
\setcounter{figure}{0}
\renewcommand{\thefigure}{A\arabic{figure}}\

\begin{appendix}

\section{unconventional optomechanical Hamiltonian}

We now present detailed derivations of unconventional cavity
optomechanics (UOM) mediated by a qubit. We start our discussions
by first considering a mechanical oscillator interacting with a
qubit with strength $g_{x}$, i,e.,
\begin{equation}
H_{\text{qm}}=\frac{1}{2}\omega_{q}\sigma_{z}+\omega_{m}b^{\dagger}b+g_{x}\sigma_{x}(b^{\dagger}+b),
\label{Hqm}
\end{equation}
where $b$ ($b^{\dagger}$) are the annihilation (creation)
operators of the mechanical mode, $\omega_{q}$ is the qubit
transition frequency, while $\sigma_{z}=|e\rangle \langle
e|-|g\rangle \langle g|$ and $\sigma_{x}=|e\rangle \langle
g|+|g\rangle \langle e|$ are the qubit Pauli operators with
$|e\rangle$ ($|g\rangle$) being the excited (ground) state. We
assume that the system is largely detuned with
$(\omega_{q}-\omega_{m})\gg g_{x}$. The optical cavity is involved
in this bipartite system by considering its longitudinal coupling
with the qubit~\cite{Liu2014,Zhao15,Richer16,Richer17}, which is
described by the Hamiltonian
\begin{equation}
H_{\text{qc}}=g_{z0}\cos(\omega_{d}t)\sigma_{z}\left(a^{\dagger}+a\right),
\label{Hqc}
\end{equation}
where $a$ ($a^{\dagger}$) are the annihilation (creation)
operators of the optical mode. As in Ref.~\cite{Didier15}, we
consider a general case, where the \emph{longitudinal coupling}
$g_{z0}$ is \emph{parametrically modulated} at a frequency
$\omega_{d}$. Note that the following discussions can also be
applied to the constant longitudinal case. Consequently, the
system Hamiltonian becomes
\begin{eqnarray}
H_{0}&=&\frac{1}{2}\omega_{q}\sigma_{z}+\omega_{c0}a^{\dagger}a+\omega_{m}b^{\dagger}b \notag \\
&&+g_{z0}\cos(\omega_{d}t)\sigma_{z}\left(a^{\dagger}+a\right)
+g_{x}\sigma_{x}(b^{\dagger}+b), \label{H0}
\end{eqnarray}
where $\omega_{c0}$ is the resonator frequency. By rotating the
resonator at frequency $\omega_{d}$, and neglecting the rapidly
oscillating terms, we obtain
\begin{equation}
H_{1}=\frac{\omega_{q}}{2}\sigma_{z}+\omega_{c}a^{\dagger}a+\omega_{m}b^{\dagger}b
+g_{z}\sigma_{z}\left(a^{\dagger}+a\right)
+g_{x}\sigma_{x}(b^{\dagger}+b), \label{H1}
\end{equation}
where $\omega_{c}=\omega_{c0}-\omega_{d}$ is the shifted resonator
frequency, and $g_{z}=g_{z0}/2$ is an effective longitudinal
coupling strength. Later we find that this modulation allows us to
obtain an exact analogue of the conventional optomechanical
Hamiltonian by dropping the quadratic terms.

By setting $\xi=(a^{\dagger}+a)$, the longitudinal interaction can
be viewed as the quantized optical field modulating the qubit
transition frequency as $$\omega_{q}(\xi)=\omega_{q}+2g_{z}\xi.$$
Here we assume that the qubit-mechanical interaction is in the
dispersive regime. By defining
$$X_{\pm}=\sigma_{-}b^{\dagger}\pm\sigma_{+}b, \quad  Y_{\pm}=\sigma_{+}b^{\dagger}\pm\sigma_{-}b,$$
we can rewrite
$H_{1}$ as
\begin{equation}
H_{2}=\frac{1}{2}\omega_{q}(\xi)\sigma_{z}+\omega_{c}a^{\dagger}a+\omega_{m}b^{\dagger}b+g_{x}\left(X_{+}+Y_{+}\right).
\label{H2}
\end{equation}
Different from the standard derivations of dispersive coupling
under the rotating wave approximation, we also consider the
counter-rotating term $Y_{+}$ in $H_{2}$. Applying the unitary
transformation~\cite{Zueco09},
\begin{gather}
U=\exp\left[\lambda_{-}(\xi) X_{-}+ \lambda_{+}(\xi) Y_{-}\right],
\quad \lambda_{\pm}(\xi)=\frac{g_{x}}{\omega_{q}(\xi)\pm
\omega_{c}},
\end{gather}
to $H_{2}$, we can expand the transformed Hamiltonian
$\widetilde{H}=U^{\dagger}H_{2}U$ to first order in the small
parameters $\lambda_{\pm}(\xi)$. Thus, we obtain the following
dispersive-type coupling Hamiltonian
\begin{eqnarray}
\widetilde{H}&\simeq& H_{0}+H_{\text{dis}}, \label{eqHt} \\
H_{0}(\xi)&=&\frac{\omega_{q}(\xi)}{2}\sigma_{z}+\omega_{c}a^{\dagger}a+\omega_{m}b^{\dagger}b,             \label{eqH0} \\
H_{\text{dis}}(\xi)&=&\frac{1}{2}\sigma_{z}
g_{x}\left[\lambda_{+}(\xi)+\lambda_{-}(\xi)\right](b^{\dagger}+b)^{2}.
\label{eqHint}
\end{eqnarray}
Comparing with the standard dispersive coupling, we find two
differences in Eq.~(\ref{eqHint}): First, due to counter-rotating
contributions, the quadratic terms $b^{2}$ and $b^{\dagger 2}$ are
also involved. Second, more importantly, the dispersive coupling
strength,
\begin{equation}
\chi(\xi)=\frac{1}{2}
g_{x}\left[\lambda_{+}(\xi)+\lambda_{-}(\xi)\right]=\frac{g_{x}^{2}\:\omega_{q}(\xi)}{\omega_{q}^{2}(\xi)-\omega_{m}^{2}},
\label{chi1}
\end{equation}
is not constant but depends on the quantized optical field
operator $\xi=a+a^{\dagger}$. Assuming that $\omega_{q}\gg
2g_{z}\xi$ and $\omega_{q}\gg \omega_{m}$, we approximately expand
$H_{\text{dis}}(\xi)$ to second order in $\xi$, and obtain
Eq.~(\ref{chiUOM}).

Given that $$\delta_{s}=(\omega_{c}-2\omega_{m}) \gg G_{1},$$ the
interaction Hamiltonian in Eq.~(\ref{Hs}) reads
$$H_{\delta}=G_{1}\left[a^{\dagger}b^{2}\exp(i\delta_{s}t)+ab^{\dagger2}\exp(-i\delta_{s}t)\right],$$ from which we can obtain the cross-Kerr interaction between these two modes~\cite{Ding17}, i.e.,
\begin{equation}
H_{\text{ck}}=\frac{2G_{1}^{2}}{\delta_{s}}\left[a^{\dagger}a(2b^{\dagger}b+1)-b^{\dagger2}b^{2}\right].
\end{equation}
This interaction describes that the average phonon number operator
$\langle n \rangle=\langle  b^{\dagger}b \rangle$, will
effectively shift the optical frequency. Equation~(\ref{Hchix})
also contains a cross-Kerr coupling with strength $-2G_{2}$ (a
second-order term). Therefore, the phonon number $\langle
n\rangle=\langle b^{\dagger}b\rangle$ eventually shifts the
optical frequency by the amount
\begin{equation}
\delta \omega_{c}=\chi_{k} \langle n\rangle,  \quad
\chi_{k}=\left[\frac{4G_{1}^{2}}{\delta_{s}}-2G_{2}\right].
\end{equation}
By setting $\delta_{s}=20G_{1}$ and adopting the parameters
specified in the main article, the optical frequency shift per
phonon is about $\chi_{k}/(2\pi)\backsimeq-10~\text{kHz}$. Since
$H_{\text{ck}}$ commutes with the phonon-number operator, this
interaction can be employed for phonon quantum nondemolition (QND)
measurements and phonon distribution
counting~\cite{Munro05,Ding17}.

Finally, we discuss the parameter regimes where this
qubit-mediated coupling for unconventional cavity optomechanics
(UOM) is valid. First, we recall that the effective interaction is
based on the dispersive coupling between the mechanical mode and
the qubit, which sets a limitation on the average phonon number as~\cite{Gu2017}
\begin{equation}
\langle n\rangle=\langle b^{\dagger}b\rangle\leq \frac{1}{(2\lambda_{-})^{2}}.
\end{equation}
 Second, the convergence of the
expansion, given in Eq.~(\ref{chiz}), requires that $\langle \xi
\rangle \leq 2g_{z}/\omega_{q}$, which also sets a bound for the
optical intracavity field amplitude.

\setcounter{equation}{0}
\renewcommand{\theequation}{B\arabic{equation}}
\setcounter{figure}{0}
\renewcommand{\thefigure}{B\arabic{figure}}\

\section{mechanical parametric amplifier based on unconventional optomechanics}

In Fig.~\ref{fig1m}(b), the TLR can be replaced by a 1D microwave
guide, which allows for a classical current signal to propagate
inside. As a result, the shape of the classical drive applied to
the qubit directly determines the boundary condition of the SAW
resonator. Here we consider the UOM system working as a
phase-sensitive mechanical parametric amplifier
(MPA)~\cite{Rugar91,Carr2000}, which can enhance one quadrature of
an ultraweak mechanical signal for quantum detection. To this end,
we assume that the longitudinal drive of the qubit has the form
$$H_{d}=\Omega_{d} \sigma_{z}\cos(2\omega_{m}t+\phi).$$ Following
the derivation steps in Sec.~I, the effective Hamiltonian reads
\begin{gather}
H_{\text{MPA}}=\alpha b^{\dagger2}+\alpha^{\ast} b^{2}, \\
\alpha=\Omega_{d}\frac{g_{x}^{2}(\omega_{q}^{2}+\omega_{m}^{2})}{(\omega_{q}^{2}-\omega_{m}^{2})^{2}}e^{-i\phi}\simeq
\Omega_{d}\left(\frac{g_{x}}{\omega_{q}}\right)^{2}e^{-i\phi},
\label{Hmpa}
\end{gather}

Since the qubit is a highly nonlinear system, it introduces a Kerr
nonlinearity in the SAW resonator~\cite{Boissonneault09}, i.e.,
$$H_{k}=K(b^{\dagger}b^{\dagger}bb)\sigma_{z}, \quad  K=\frac{g_{x}^{4}}{\omega_{q}^{3}}.$$
We consider that a weak mechanical
signal is injected into the SAW resonator, i.e.,
$$H_{\text{in}}=i\sqrt{\kappa}(c_{\text{in}}^{\dagger}b-c_{\text{in}}b^{\dagger}),$$
where $\kappa$ is the phonon damping rate of the input channel,
and $c_{\text{in}}$ is the input field operator. The output signal
$c_{\text{out}}$ can be obtained via the input-output relation
$$c_{\text{out}}(t)=\sqrt{\kappa}b(t)+c_{\text{in}}.$$ The
Heisenberg equation for the phonon operator $b(t)$ reads
\begin{equation}
\dot{b}(t)=-2i\alpha
b^{\dagger}(t)-\frac{\kappa}{2}b(t)-2iKN(t)\sigma_{z}(t)b(t)-\sqrt{\kappa}c_{\text{in}},
\label{bH}
\end{equation}
where $N(t)=\langle b^{\dagger}(t)b(t)\rangle$ is the mean phonon
number. We define the input (output) quadratures
$$X_{\text{in,out}}=\frac{c_{\text{in,out}}^{\dagger}+c_{\text{in,out}}}{\sqrt{2}},
\quad Y_{\text{in,out}}=\frac{i(c_{\text{in,out}}^{\dagger}-c_{\text{in,out}})}{\sqrt{2}}.$$

As we have discussed in Sec.~I, the qubit is approximately in its
ground state. Therefore, we set $\sigma_{z}=-1$ and obtain
\begin{eqnarray}
\dot{X}(t)&=&-\frac{1}{2}\kappa X(t)-2KN(t)Y(t)  \notag \\
&&-2|\alpha|[\cos{\phi}Y(t)+\sin{\phi}X(t)]-\sqrt{\kappa}X_{\text{in}}, \\
\dot{Y}(t)&=&-\frac{1}{2}\kappa Y(t)+2KN(t)X(t) \notag \\
&&+2|\alpha|[\cos{\phi}X(t)+\sin{\phi}Y(t)]-\sqrt{\kappa}Y_{\text{in}},
\label{XY}
\end{eqnarray}
where $$X=\frac{b^{\dagger}+b}{\sqrt{2}}, \quad  Y=\frac{i(b^{\dagger}-b)}{\sqrt{2}}$$ are the intracavity quadratures. We
assume that the quadrature $X_{\text{in}}$ is to be amplified and
satisfies the boundary relation
$X_{\text{out}}=\sqrt{\kappa}X+X_{\text{in}}$. From Eq.~(\ref{XY})
one can find that, only under the conditions $\cos{\phi}=0$ and
$K=0$, the evolutions of the quadratures $X$ and $Y$ are
decoupled, and one can amplify $X_{\text{in}}$
independently~\cite{Clerk10}. The Kerr nonlinearity couples both
quadratures and should be avoided. In experiments, $K$ cannot be
exactly equal to zero. To minimize the effects of the quadrature
$Y$ on $X$, we require that the induced Kerr nonlinearity term
satisfies $KN(t)\ll |\alpha|$, which leads to
\begin{equation}
N\ll
\frac{|\alpha|}{K}\simeq\frac{\Omega_{d}\omega_{q}}{g_{x}^{2}}.
\label{N11}
\end{equation}
Therefore, the qubit-induced Kerr nonlinearity sets an upper bound
of the mean phonon number, below which one can safely neglect the
Kerr terms in Eq.~(\ref{XY}). We consider that the amplification
process works below the threshold regime
($2|\alpha|<\kappa/2$)~\cite{Scully1997}, and the system reaches
its steady state when $t\rightarrow\infty$. At $\phi=-\pi/2$, the
quadrature $X_{\text{in}}$ ($Y_{\text{in}}$) is amplified
(attenuated) as
\begin{subequations}
    \begin{gather}
    \frac{\langle X_{\text{out}}\rangle }{\langle X_{\text{in}}\rangle}=G, \qquad \frac{\langle Y_{\text{out}}\rangle }{\langle Y_{\text{in}}\rangle}=\frac{1}{G}. \\
    G=\left|\frac{4|\alpha|+\kappa}{4|\alpha|-\kappa}\right|.
    \end{gather}
\end{subequations}
Therefore, the dynamics in Eq.~(\ref{XY}) causes one signal
quadrature of the input field to be amplified while the conjugate
one to be attenuated. Since the commutation relation is preserved,
no extra noise is introduced in principle~\cite{Caves1982}. In the
numerical calculations in Fig.~\ref{fig3m} in the main article, we set
$\langle c_{\text{in}}\rangle$ as a real number. Moreover, if we
consider that $\phi$ is shifted away from $\pi/2$, the gain now
becomes dependent on the phase $\phi$, which is shown in
Eq.~(\ref{gainG}). We have assumed that the input signal is much
weaker compared with the amplified strength, i.e.,
$\sqrt{\kappa}\langle c_{\text{in}} \rangle \ll |\alpha|$. In the
steady state, the phonon number inside the SAW resonator
is~\cite{Scully1997}
\begin{equation}
N=\langle
b^{\dagger}b\rangle=\frac{8|\alpha|^{2}}{\kappa^{2}-16|\alpha|^{2}}.
\end{equation}

In an ideal MPA case, $4|\alpha|$ can approach $\kappa$ with a
very small deviation. As a result, the steady-state phonon number
can be ultra-large when $4|\alpha|\simeq \kappa$. However, our
proposal is based on an UOM system mediated by a qubit. As shown
in Eq.~(\ref{B9}), during the amplifying process, the
qubit-induced Kerr nonlinearity sets a limitation on the phonon
number. Moreover, when deriving Hamiltonian (\ref{Hmpa}), we have
assumed that the qubit is approximately in its ground state. When
$N$ is too large, the qubit can be effectively excited. The
derivation given in Sec.~I is not valid any more. We should make
sure that our proposal is in the quasi-dispersive
regime~\cite{Boissonneault09}, i.e.,
$N<\omega_{q}^{2}/(4g_{x}^{2})$. As a result, the critical phonon
number should satisfy
\begin{equation}
N_{c}=\max\left\{\frac{\Omega_{d}\omega_{q}}{g_{x}^{2}},
\frac{\omega_{q}^{2}}{4g_{x}^{2}}\right\}.
\label{B9}
\end{equation}
The UOM system can work as an effective MPA below the critical
phonon number $N_{c}$. Therefore, the amplitude $|\alpha|$ should
satisfy $$4|\alpha|<\kappa\sqrt{2N_{c}/(2N_{c}+1)},$$ which
consequently leads to a critical maximum gain in
Eq.~(\ref{gainmax}), which is valid when $N_{c}\gg1$. Thus, the maximum
mechanical amplification gain is bounded by $G_{\text{c}}$.
\end{appendix}


\begin{thebibliography}{77}%
	\makeatletter
	\providecommand \@ifxundefined [1]{%
		\@ifx{#1\undefined}
	}%
	\providecommand \@ifnum [1]{%
		\ifnum #1\expandafter \@firstoftwo
		\else \expandafter \@secondoftwo
		\fi
	}%
	\providecommand \@ifx [1]{%
		\ifx #1\expandafter \@firstoftwo
		\else \expandafter \@secondoftwo
		\fi
	}%
	\providecommand \natexlab [1]{#1}%
	\providecommand \enquote  [1]{``#1''}%
	\providecommand \bibnamefont  [1]{#1}%
	\providecommand \bibfnamefont [1]{#1}%
	\providecommand \citenamefont [1]{#1}%
	\providecommand \href@noop [0]{\@secondoftwo}%
	\providecommand \href [0]{\begingroup \@sanitize@url \@href}%
	\providecommand \@href[1]{\@@startlink{#1}\@@href}%
	\providecommand \@@href[1]{\endgroup#1\@@endlink}%
	\providecommand \@sanitize@url [0]{\catcode `\\12\catcode `\$12\catcode
		`\&12\catcode `\#12\catcode `\^12\catcode `\_12\catcode `\%12\relax}%
	\providecommand \@@startlink[1]{}%
	\providecommand \@@endlink[0]{}%
	\providecommand \url  [0]{\begingroup\@sanitize@url \@url }%
	\providecommand \@url [1]{\endgroup\@href {#1}{\urlprefix }}%
	\providecommand \urlprefix  [0]{URL }%
	\providecommand \Eprint [0]{\href }%
	\providecommand \doibase [0]{http://dx.doi.org/}%
	\providecommand \selectlanguage [0]{\@gobble}%
	\providecommand \bibinfo  [0]{\@secondoftwo}%
	\providecommand \bibfield  [0]{\@secondoftwo}%
	\providecommand \translation [1]{[#1]}%
	\providecommand \BibitemOpen [0]{}%
	\providecommand \bibitemStop [0]{}%
	\providecommand \bibitemNoStop [0]{.\EOS\space}%
	\providecommand \EOS [0]{\spacefactor3000\relax}%
	\providecommand \BibitemShut  [1]{\csname bibitem#1\endcsname}%
	\let\auto@bib@innerbib\@empty
	\bibitem [{\citenamefont {Bowen}\ and\ \citenamefont
		{Milburn}(2015)}]{bowen2015quantum}%
	\BibitemOpen
	\bibfield  {author} {\bibinfo {author} {\bibfnamefont {W.~P.}\ \bibnamefont
			{Bowen}}\ and\ \bibinfo {author} {\bibfnamefont {G.~J.}\ \bibnamefont
			{Milburn}},\ }\href@noop {} {\emph {\bibinfo {title} {Quantum
				optomechanics}}}\ (\bibinfo  {publisher} {CRC press},\ \bibinfo {year}
	{2015})\BibitemShut {NoStop}%
	\bibitem [{\citenamefont {Aspelmeyer}\ \emph
		{et~al.}(2014{\natexlab{a}})\citenamefont {Aspelmeyer}, \citenamefont
		{Kippenberg},\ and\ \citenamefont {Marquardt}}]{aspelmeyer2014cavity}%
	\BibitemOpen
	\bibfield  {author} {\bibinfo {author} {\bibfnamefont {M.}~\bibnamefont
			{Aspelmeyer}}, \bibinfo {author} {\bibfnamefont {T.~J.}\ \bibnamefont
			{Kippenberg}}, \ and\ \bibinfo {author} {\bibfnamefont {F.}~\bibnamefont
			{Marquardt}},\ }\href@noop {} {\emph {\bibinfo {title} {Cavity optomechanics:
				nano- and micromechanical resonators interacting with light}}}\ (\bibinfo
	{publisher} {Springer},\ \bibinfo {year} {2014})\BibitemShut {NoStop}%
	\bibitem [{\citenamefont {Law}(1995)}]{Law95}%
	\BibitemOpen
	\bibfield  {author} {\bibinfo {author} {\bibfnamefont {C.~K.}\ \bibnamefont
			{Law}},\ }\bibfield  {title} {\enquote {\bibinfo {title} {Interaction between
				a moving mirror and radiation pressure: A {H}amiltonian formulation},}\
	}\href {\doibase 10.1103/PhysRevA.51.2537} {\bibfield  {journal} {\bibinfo
			{journal} {Phys. Rev. A}\ }\textbf {\bibinfo {volume} {51}},\ \bibinfo
		{pages} {2537} (\bibinfo {year} {1995})}\BibitemShut {NoStop}%
	\bibitem [{\citenamefont {Bose}\ \emph {et~al.}(1997)\citenamefont {Bose},
		\citenamefont {Jacobs},\ and\ \citenamefont {Knight}}]{Bose97}%
	\BibitemOpen
	\bibfield  {author} {\bibinfo {author} {\bibfnamefont {S.}~\bibnamefont
			{Bose}}, \bibinfo {author} {\bibfnamefont {K.}~\bibnamefont {Jacobs}}, \ and\
		\bibinfo {author} {\bibfnamefont {P.~L.}\ \bibnamefont {Knight}},\ }\bibfield
	{title} {\enquote {\bibinfo {title} {Preparation of nonclassical states in
				cavities with a moving mirror},}\ }\href {\doibase 10.1103/PhysRevA.56.4175}
	{\bibfield  {journal} {\bibinfo  {journal} {Phys. Rev. A}\ }\textbf {\bibinfo
			{volume} {56}},\ \bibinfo {pages} {4175} (\bibinfo {year}
		{1997})}\BibitemShut {NoStop}%
	\bibitem [{\citenamefont {Bassi}\ \emph {et~al.}(2013)\citenamefont {Bassi},
		\citenamefont {Lochan}, \citenamefont {Satin}, \citenamefont {Singh},\ and\
		\citenamefont {Ulbricht}}]{Bassi13}%
	\BibitemOpen
	\bibfield  {author} {\bibinfo {author} {\bibfnamefont {A.}~\bibnamefont
			{Bassi}}, \bibinfo {author} {\bibfnamefont {K.}~\bibnamefont {Lochan}},
		\bibinfo {author} {\bibfnamefont {S.}~\bibnamefont {Satin}}, \bibinfo
		{author} {\bibfnamefont {T.~P.}\ \bibnamefont {Singh}}, \ and\ \bibinfo
		{author} {\bibfnamefont {H.}~\bibnamefont {Ulbricht}},\ }\bibfield  {title}
	{\enquote {\bibinfo {title} {Models of wave-function collapse, underlying
				theories, and experimental tests},}\ }\href {\doibase
		10.1103/RevModPhys.85.471} {\bibfield  {journal} {\bibinfo  {journal} {Rev.
				Mod. Phys.}\ }\textbf {\bibinfo {volume} {85}},\ \bibinfo {pages} {471}
		(\bibinfo {year} {2013})}\BibitemShut {NoStop}%
	\bibitem [{\citenamefont {Blencowe}(2013)}]{Blencowe13}%
	\BibitemOpen
	\bibfield  {author} {\bibinfo {author} {\bibfnamefont {M.~P.}\ \bibnamefont
			{Blencowe}},\ }\bibfield  {title} {\enquote {\bibinfo {title} {Effective
				field theory approach to gravitationally induced decoherence},}\ }\href
	{\doibase 10.1103/PhysRevLett.111.021302} {\bibfield  {journal} {\bibinfo
			{journal} {Phys. Rev. Lett.}\ }\textbf {\bibinfo {volume} {111}},\ \bibinfo
		{pages} {021302} (\bibinfo {year} {2013})}\BibitemShut {NoStop}%
	\bibitem [{\citenamefont {Schaller}\ \emph {et~al.}(2002)\citenamefont
		{Schaller}, \citenamefont {Sch\"utzhold}, \citenamefont {Plunien},\ and\
		\citenamefont {Soff}}]{Schaller02}%
	\BibitemOpen
	\bibfield  {author} {\bibinfo {author} {\bibfnamefont {G.}~\bibnamefont
			{Schaller}}, \bibinfo {author} {\bibfnamefont {R.}~\bibnamefont
			{Sch\"utzhold}}, \bibinfo {author} {\bibfnamefont {G.}~\bibnamefont
			{Plunien}}, \ and\ \bibinfo {author} {\bibfnamefont {G.}~\bibnamefont
			{Soff}},\ }\bibfield  {title} {\enquote {\bibinfo {title} {Dynamical
				{C}asimir effect in a leaky cavity at finite temperature},}\ }\href {\doibase
		10.1103/PhysRevA.66.023812} {\bibfield  {journal} {\bibinfo  {journal} {Phys.
				Rev. A}\ }\textbf {\bibinfo {volume} {66}},\ \bibinfo {pages} {023812}
		(\bibinfo {year} {2002})}\BibitemShut {NoStop}%
	\bibitem [{\citenamefont {Kim}\ \emph {et~al.}(2006)\citenamefont {Kim},
		\citenamefont {Brownell},\ and\ \citenamefont {Onofrio}}]{Kim06}%
	\BibitemOpen
	\bibfield  {author} {\bibinfo {author} {\bibfnamefont {W.-J.}\ \bibnamefont
			{Kim}}, \bibinfo {author} {\bibfnamefont {J.~H.}\ \bibnamefont {Brownell}}, \
		and\ \bibinfo {author} {\bibfnamefont {R.}~\bibnamefont {Onofrio}},\
	}\bibfield  {title} {\enquote {\bibinfo {title} {Detectability of dissipative
				motion in quantum vacuum via superradiance},}\ }\href {\doibase
		10.1103/PhysRevLett.96.200402} {\bibfield  {journal} {\bibinfo  {journal}
			{Phys. Rev. Lett.}\ }\textbf {\bibinfo {volume} {96}},\ \bibinfo {pages}
		{200402} (\bibinfo {year} {2006})}\BibitemShut {NoStop}%
	\bibitem [{\citenamefont {Johansson}\ \emph {et~al.}(2009)\citenamefont
		{Johansson}, \citenamefont {Johansson}, \citenamefont {Wilson},\ and\
		\citenamefont {Nori}}]{Johansson09L}%
	\BibitemOpen
	\bibfield  {author} {\bibinfo {author} {\bibfnamefont {J.~R.}\ \bibnamefont
			{Johansson}}, \bibinfo {author} {\bibfnamefont {G.}~\bibnamefont
			{Johansson}}, \bibinfo {author} {\bibfnamefont {C.~M.}\ \bibnamefont
			{Wilson}}, \ and\ \bibinfo {author} {\bibfnamefont {F.}~\bibnamefont
			{Nori}},\ }\bibfield  {title} {\enquote {\bibinfo {title} {Dynamical
				{C}asimir effect in a superconducting coplanar waveguide},}\ }\href {\doibase
		10.1103/PhysRevLett.103.147003} {\bibfield  {journal} {\bibinfo  {journal}
			{Phys. Rev. Lett.}\ }\textbf {\bibinfo {volume} {103}},\ \bibinfo {pages}
		{147003} (\bibinfo {year} {2009})}\BibitemShut {NoStop}%
	\bibitem [{\citenamefont {Wilson}\ \emph {et~al.}(2011)\citenamefont {Wilson},
		\citenamefont {Johansson}, \citenamefont {Pourkabirian}, \citenamefont
		{Simoen}, \citenamefont {Johansson}, \citenamefont {Duty}, \citenamefont
		{Nori},\ and\ \citenamefont {Delsing}}]{Wilson2011}%
	\BibitemOpen
	\bibfield  {author} {\bibinfo {author} {\bibfnamefont {C.~M.}\ \bibnamefont
			{Wilson}}, \bibinfo {author} {\bibfnamefont {G.}~\bibnamefont {Johansson}},
		\bibinfo {author} {\bibfnamefont {A.}~\bibnamefont {Pourkabirian}}, \bibinfo
		{author} {\bibfnamefont {M.}~\bibnamefont {Simoen}}, \bibinfo {author}
		{\bibfnamefont {J.~R.}\ \bibnamefont {Johansson}}, \bibinfo {author}
		{\bibfnamefont {T.}~\bibnamefont {Duty}}, \bibinfo {author} {\bibfnamefont
			{F.}~\bibnamefont {Nori}}, \ and\ \bibinfo {author} {\bibfnamefont
			{P.}~\bibnamefont {Delsing}},\ }\bibfield  {title} {\enquote {\bibinfo
			{title} {Observation of the dynamical {C}asimir effect in a superconducting
				circuit},}\ }\href {\doibase 10.1038/nature10561} {\bibfield  {journal}
		{\bibinfo  {journal} {Nature (London)}\ }\textbf {\bibinfo {volume} {479}},\
		\bibinfo {pages} {376} (\bibinfo {year} {2011})}\BibitemShut {NoStop}%
	\bibitem [{\citenamefont {Nation}\ \emph {et~al.}(2012)\citenamefont {Nation},
		\citenamefont {Johansson}, \citenamefont {Blencowe},\ and\ \citenamefont
		{Nori}}]{Nation12}%
	\BibitemOpen
	\bibfield  {author} {\bibinfo {author} {\bibfnamefont {P.~D.}\ \bibnamefont
			{Nation}}, \bibinfo {author} {\bibfnamefont {J.~R.}\ \bibnamefont
			{Johansson}}, \bibinfo {author} {\bibfnamefont {M.~P.}\ \bibnamefont
			{Blencowe}}, \ and\ \bibinfo {author} {\bibfnamefont {F.}~\bibnamefont
			{Nori}},\ }\bibfield  {title} {\enquote {\bibinfo {title} {Colloquium:
				Stimulating uncertainty: Amplifying the quantum vacuum with superconducting
				circuits},}\ }\href {\doibase 10.1103/RevModPhys.84.1} {\bibfield  {journal}
		{\bibinfo  {journal} {Rev. Mod. Phys.}\ }\textbf {\bibinfo {volume} {84}},\
		\bibinfo {pages} {1} (\bibinfo {year} {2012})}\BibitemShut {NoStop}%
	\bibitem [{\citenamefont {Johansson}\ \emph
		{et~al.}(2013{\natexlab{a}})\citenamefont {Johansson}, \citenamefont
		{Johansson}, \citenamefont {Wilson}, \citenamefont {Delsing},\ and\
		\citenamefont {Nori}}]{Johansson13}%
	\BibitemOpen
	\bibfield  {author} {\bibinfo {author} {\bibfnamefont {J.~R.}\ \bibnamefont
			{Johansson}}, \bibinfo {author} {\bibfnamefont {G.}~\bibnamefont
			{Johansson}}, \bibinfo {author} {\bibfnamefont {C.~M.}\ \bibnamefont
			{Wilson}}, \bibinfo {author} {\bibfnamefont {P.}~\bibnamefont {Delsing}}, \
		and\ \bibinfo {author} {\bibfnamefont {F.}~\bibnamefont {Nori}},\ }\bibfield
	{title} {\enquote {\bibinfo {title} {Nonclassical microwave radiation from
				the dynamical {C}asimir effect},}\ }\href {\doibase
		10.1103/PhysRevA.87.043804} {\bibfield  {journal} {\bibinfo  {journal} {Phys.
				Rev. A}\ }\textbf {\bibinfo {volume} {87}},\ \bibinfo {pages} {043804}
		(\bibinfo {year} {2013}{\natexlab{a}})}\BibitemShut {NoStop}%
	\bibitem [{\citenamefont {Macr\`{\i}}\ \emph {et~al.}(2018)\citenamefont
		{Macr\`{\i}}, \citenamefont {Ridolfo}, \citenamefont {Di~Stefano},
		\citenamefont {Kockum}, \citenamefont {Nori},\ and\ \citenamefont
		{Savasta}}]{Macr2018}%
	\BibitemOpen
	\bibfield  {author} {\bibinfo {author} {\bibfnamefont {V.}~\bibnamefont
			{Macr\`{\i}}}, \bibinfo {author} {\bibfnamefont {A.}~\bibnamefont {Ridolfo}},
		\bibinfo {author} {\bibfnamefont {O.}~\bibnamefont {Di~Stefano}}, \bibinfo
		{author} {\bibfnamefont {A.~F.}\ \bibnamefont {Kockum}}, \bibinfo {author}
		{\bibfnamefont {F.}~\bibnamefont {Nori}}, \ and\ \bibinfo {author}
		{\bibfnamefont {S.}~\bibnamefont {Savasta}},\ }\bibfield  {title} {\enquote
		{\bibinfo {title} {Nonperturbative dynamical {C}asimir effect in
				optomechanical systems: Vacuum {C}asimir-{R}abi splittings},}\ }\href
	{\doibase 10.1103/PhysRevX.8.011031} {\bibfield  {journal} {\bibinfo
			{journal} {Phys. Rev. X}\ }\textbf {\bibinfo {volume} {8}},\ \bibinfo {pages}
		{011031} (\bibinfo {year} {2018})}\BibitemShut {NoStop}%
	\bibitem [{\citenamefont {Di~Stefano}\ \emph {et~al.}(2019)\citenamefont
		{Di~Stefano}, \citenamefont {Settineri}, \citenamefont {Macr\`{\i}},
		\citenamefont {Ridolfo}, \citenamefont {Stassi}, \citenamefont {Kockum},
		\citenamefont {Savasta},\ and\ \citenamefont {Nori}}]{Stefano18a}%
	\BibitemOpen
	\bibfield  {author} {\bibinfo {author} {\bibfnamefont {O.}~\bibnamefont
			{Di~Stefano}}, \bibinfo {author} {\bibfnamefont {A.}~\bibnamefont
			{Settineri}}, \bibinfo {author} {\bibfnamefont {V.}~\bibnamefont
			{Macr\`{\i}}}, \bibinfo {author} {\bibfnamefont {A.}~\bibnamefont {Ridolfo}},
		\bibinfo {author} {\bibfnamefont {R.}~\bibnamefont {Stassi}}, \bibinfo
		{author} {\bibfnamefont {A.~F.}\ \bibnamefont {Kockum}}, \bibinfo {author}
		{\bibfnamefont {S.}~\bibnamefont {Savasta}}, \ and\ \bibinfo {author}
		{\bibfnamefont {F.}~\bibnamefont {Nori}},\ }\bibfield  {title} {\enquote
		{\bibinfo {title} {Interaction of mechanical oscillators mediated by the
				exchange of virtual photon pairs},}\ }\href {\doibase
		10.1103/PhysRevLett.122.030402} {\bibfield  {journal} {\bibinfo  {journal}
			{Phys. Rev. Lett.}\ }\textbf {\bibinfo {volume} {122}},\ \bibinfo {pages}
		{03002} (\bibinfo {year} {2019})}\BibitemShut {NoStop}%
	\bibitem [{\citenamefont {Marshall}\ \emph {et~al.}(2003)\citenamefont
		{Marshall}, \citenamefont {Simon}, \citenamefont {Penrose},\ and\
		\citenamefont {Bouwmeester}}]{Marshall03}%
	\BibitemOpen
	\bibfield  {author} {\bibinfo {author} {\bibfnamefont {W.}~\bibnamefont
			{Marshall}}, \bibinfo {author} {\bibfnamefont {C.}~\bibnamefont {Simon}},
		\bibinfo {author} {\bibfnamefont {R.}~\bibnamefont {Penrose}}, \ and\
		\bibinfo {author} {\bibfnamefont {D.}~\bibnamefont {Bouwmeester}},\
	}\bibfield  {title} {\enquote {\bibinfo {title} {Towards quantum
				superpositions of a mirror},}\ }\href {\doibase
		10.1103/PhysRevLett.91.130401} {\bibfield  {journal} {\bibinfo  {journal}
			{Phys. Rev. Lett.}\ }\textbf {\bibinfo {volume} {91}},\ \bibinfo {pages}
		{130401} (\bibinfo {year} {2003})}\BibitemShut {NoStop}%
	\bibitem [{\citenamefont {Liao}\ and\ \citenamefont {Tian}(2016)}]{Liao16}%
	\BibitemOpen
	\bibfield  {author} {\bibinfo {author} {\bibfnamefont {J.-Q.}\ \bibnamefont
			{Liao}}\ and\ \bibinfo {author} {\bibfnamefont {L.}~\bibnamefont {Tian}},\
	}\bibfield  {title} {\enquote {\bibinfo {title} {Macroscopic quantum
				superposition in cavity optomechanics},}\ }\href {\doibase
		10.1103/PhysRevLett.116.163602} {\bibfield  {journal} {\bibinfo  {journal}
			{Phys. Rev. Lett.}\ }\textbf {\bibinfo {volume} {116}},\ \bibinfo {pages}
		{163602} (\bibinfo {year} {2016})}\BibitemShut {NoStop}%
	\bibitem [{\citenamefont {Carusotto}\ \emph {et~al.}(2009)\citenamefont
		{Carusotto}, \citenamefont {Balbinot}, \citenamefont {Fabbri},\ and\
		\citenamefont {Recati}}]{Carusotto2009}%
	\BibitemOpen
	\bibfield  {author} {\bibinfo {author} {\bibfnamefont {I.}~\bibnamefont
			{Carusotto}}, \bibinfo {author} {\bibfnamefont {R.}~\bibnamefont {Balbinot}},
		\bibinfo {author} {\bibfnamefont {A.}~\bibnamefont {Fabbri}}, \ and\ \bibinfo
		{author} {\bibfnamefont {A.}~\bibnamefont {Recati}},\ }\bibfield  {title}
	{\enquote {\bibinfo {title} {Density correlations and analog dynamical
				{C}asimir emission of {B}ogoliubov phonons in modulated atomic
				{B}ose-{E}instein condensates},}\ }\href {\doibase
		10.1140/epjd/e2009-00314-3} {\bibfield  {journal} {\bibinfo  {journal} {Eur.
				Phys. J. D}\ }\textbf {\bibinfo {volume} {56}},\ \bibinfo {pages} {391}
		(\bibinfo {year} {2009})}\BibitemShut {NoStop}%
	\bibitem [{\citenamefont {Boiron}\ \emph {et~al.}(2015)\citenamefont {Boiron},
		\citenamefont {Fabbri}, \citenamefont {Larr\'e}, \citenamefont {Pavloff},
		\citenamefont {Westbrook},\ and\ \citenamefont {Zi\ifmmode~\acute{n}\else
			\'{n}\fi{}}}]{Boiron15}%
	\BibitemOpen
	\bibfield  {author} {\bibinfo {author} {\bibfnamefont {D.}~\bibnamefont
			{Boiron}}, \bibinfo {author} {\bibfnamefont {A.}~\bibnamefont {Fabbri}},
		\bibinfo {author} {\bibfnamefont {P.-\'E.}\ \bibnamefont {Larr\'e}}, \bibinfo
		{author} {\bibfnamefont {N.}~\bibnamefont {Pavloff}}, \bibinfo {author}
		{\bibfnamefont {C.~I.}\ \bibnamefont {Westbrook}}, \ and\ \bibinfo {author}
		{\bibfnamefont {P.}~\bibnamefont {Zi\ifmmode~\acute{n}\else \'{n}\fi{}}},\
	}\bibfield  {title} {\enquote {\bibinfo {title} {Quantum signature of analog
				{H}awking radiation in momentum space},}\ }\href {\doibase
		10.1103/PhysRevLett.115.025301} {\bibfield  {journal} {\bibinfo  {journal}
			{Phys. Rev. Lett.}\ }\textbf {\bibinfo {volume} {115}},\ \bibinfo {pages}
		{025301} (\bibinfo {year} {2015})}\BibitemShut {NoStop}%
	\bibitem [{\citenamefont {Eckel}\ \emph {et~al.}(2018)\citenamefont {Eckel},
		\citenamefont {Kumar}, \citenamefont {Jacobson}, \citenamefont {Spielman},\
		and\ \citenamefont {Campbell}}]{Eckel18}%
	\BibitemOpen
	\bibfield  {author} {\bibinfo {author} {\bibfnamefont {S.}~\bibnamefont
			{Eckel}}, \bibinfo {author} {\bibfnamefont {A.}~\bibnamefont {Kumar}},
		\bibinfo {author} {\bibfnamefont {T.}~\bibnamefont {Jacobson}}, \bibinfo
		{author} {\bibfnamefont {I.~B.}\ \bibnamefont {Spielman}}, \ and\ \bibinfo
		{author} {\bibfnamefont {G.~K.}\ \bibnamefont {Campbell}},\ }\bibfield
	{title} {\enquote {\bibinfo {title} {A rapidly expanding {B}ose-{E}instein
				condensate: An expanding universe in the lab},}\ }\href {\doibase
		10.1103/PhysRevX.8.021021} {\bibfield  {journal} {\bibinfo  {journal} {Phys.
				Rev. X}\ }\textbf {\bibinfo {volume} {8}},\ \bibinfo {pages} {021021}
		(\bibinfo {year} {2018})}\BibitemShut {NoStop}%
	\bibitem [{\citenamefont {Schmit}\ \emph {et~al.}(2018)\citenamefont {Schmit},
		\citenamefont {Taketani},\ and\ \citenamefont {Wilhelm}}]{Schmit2018}%
	\BibitemOpen
	\bibfield  {author} {\bibinfo {author} {\bibfnamefont {R.~P.}\ \bibnamefont
			{Schmit}}, \bibinfo {author} {\bibfnamefont {B.~G.}\ \bibnamefont
			{Taketani}}, \ and\ \bibinfo {author} {\bibfnamefont {F.~K.}\ \bibnamefont
			{Wilhelm}},\ }\bibfield  {title} {\enquote {\bibinfo {title} {Quantum
				simulation of {H}awking radiation with surface acoustic waves},}\ }\href
	{https://arxiv.org/abs/1804.04092} {\bibfield  {journal} {\bibinfo  {journal}
			{preprint arXiv:1804.04092}\ } (\bibinfo {year} {2018})}\BibitemShut
	{NoStop}%
	\bibitem [{\citenamefont {Rugar}\ and\ \citenamefont
		{Gr\"utter}(1991)}]{Rugar91}%
	\BibitemOpen
	\bibfield  {author} {\bibinfo {author} {\bibfnamefont {D.}~\bibnamefont
			{Rugar}}\ and\ \bibinfo {author} {\bibfnamefont {P.}~\bibnamefont
			{Gr\"utter}},\ }\bibfield  {title} {\enquote {\bibinfo {title} {Mechanical
				parametric amplification and thermomechanical noise squeezing},}\ }\href
	{\doibase 10.1103/PhysRevLett.67.699} {\bibfield  {journal} {\bibinfo
			{journal} {Phys. Rev. Lett.}\ }\textbf {\bibinfo {volume} {67}},\ \bibinfo
		{pages} {699} (\bibinfo {year} {1991})}\BibitemShut {NoStop}%
	\bibitem [{\citenamefont {Carr}\ \emph {et~al.}(2000)\citenamefont {Carr},
		\citenamefont {Evoy}, \citenamefont {Sekaric}, \citenamefont {Craighead},\
		and\ \citenamefont {Parpia}}]{Carr2000}%
	\BibitemOpen
	\bibfield  {author} {\bibinfo {author} {\bibfnamefont {D.~W.}\ \bibnamefont
			{Carr}}, \bibinfo {author} {\bibfnamefont {S.}~\bibnamefont {Evoy}}, \bibinfo
		{author} {\bibfnamefont {L.}~\bibnamefont {Sekaric}}, \bibinfo {author}
		{\bibfnamefont {H.~G.}\ \bibnamefont {Craighead}}, \ and\ \bibinfo {author}
		{\bibfnamefont {J.~M.}\ \bibnamefont {Parpia}},\ }\bibfield  {title}
	{\enquote {\bibinfo {title} {Parametric amplification in a torsional
				microresonator},}\ }\href {\doibase 10.1063/1.1308270} {\bibfield  {journal}
		{\bibinfo  {journal} {Appl. Phys. Lett.}\ }\textbf {\bibinfo {volume} {77}},\
		\bibinfo {pages} {1545} (\bibinfo {year} {2000})}\BibitemShut {NoStop}%
	\bibitem [{\citenamefont {Zalalutdinov}\ \emph {et~al.}(2001)\citenamefont
		{Zalalutdinov}, \citenamefont {Olkhovets}, \citenamefont {Zehnder},
		\citenamefont {Ilic}, \citenamefont {Czaplewski}, \citenamefont {Craighead},\
		and\ \citenamefont {Parpia}}]{Zalalutdinov2001}%
	\BibitemOpen
	\bibfield  {author} {\bibinfo {author} {\bibfnamefont {M.}~\bibnamefont
			{Zalalutdinov}}, \bibinfo {author} {\bibfnamefont {A.}~\bibnamefont
			{Olkhovets}}, \bibinfo {author} {\bibfnamefont {A.}~\bibnamefont {Zehnder}},
		\bibinfo {author} {\bibfnamefont {B.}~\bibnamefont {Ilic}}, \bibinfo {author}
		{\bibfnamefont {D.}~\bibnamefont {Czaplewski}}, \bibinfo {author}
		{\bibfnamefont {H.~G.}\ \bibnamefont {Craighead}}, \ and\ \bibinfo {author}
		{\bibfnamefont {J.~M.}\ \bibnamefont {Parpia}},\ }\bibfield  {title}
	{\enquote {\bibinfo {title} {Optically pumped parametric amplification for
				micromechanical oscillators},}\ }\href {\doibase 10.1063/1.1371248}
	{\bibfield  {journal} {\bibinfo  {journal} {Appl. Phys. Lett.}\ }\textbf
		{\bibinfo {volume} {78}},\ \bibinfo {pages} {3142} (\bibinfo {year}
		{2001})}\BibitemShut {NoStop}%
	\bibitem [{\citenamefont {Karabalin}\ \emph {et~al.}(2010)\citenamefont
		{Karabalin}, \citenamefont {Masmanidis},\ and\ \citenamefont
		{Roukes}}]{Karabalin2010}%
	\BibitemOpen
	\bibfield  {author} {\bibinfo {author} {\bibfnamefont {R.~B.}\ \bibnamefont
			{Karabalin}}, \bibinfo {author} {\bibfnamefont {S.~C.}\ \bibnamefont
			{Masmanidis}}, \ and\ \bibinfo {author} {\bibfnamefont {M.~L.}\ \bibnamefont
			{Roukes}},\ }\bibfield  {title} {\enquote {\bibinfo {title} {Efficient
				parametric amplification in high and very high frequency piezoelectric
				nanoelectromechanical systems},}\ }\href {\doibase 10.1063/1.3505500}
	{\bibfield  {journal} {\bibinfo  {journal} {Appl. Phys. Lett.}\ }\textbf
		{\bibinfo {volume} {97}},\ \bibinfo {pages} {183101} (\bibinfo {year}
		{2010})}\BibitemShut {NoStop}%
	\bibitem [{\citenamefont {Gu}\ \emph {et~al.}(2017)\citenamefont {Gu},
		\citenamefont {Kockum}, \citenamefont {Miranowicz}, \citenamefont {Liu},\
		and\ \citenamefont {Nori}}]{Gu2017}%
	\BibitemOpen
	\bibfield  {author} {\bibinfo {author} {\bibfnamefont {X.}~\bibnamefont
			{Gu}}, \bibinfo {author} {\bibfnamefont {A.~F.}\ \bibnamefont {Kockum}},
		\bibinfo {author} {\bibfnamefont {A.}~\bibnamefont {Miranowicz}}, \bibinfo
		{author} {\bibfnamefont {Y.-X.}\ \bibnamefont {Liu}}, \ and\ \bibinfo
		{author} {\bibfnamefont {F.}~\bibnamefont {Nori}},\ }\bibfield  {title}
	{\enquote {\bibinfo {title} {Microwave photonics with superconducting quantum
				circuits},}\ }\href {https://doi.org/10.1016/j.physrep.2017.10.002}
	{\bibfield  {journal} {\bibinfo  {journal} {Phys. Rep.}\ }\textbf {\bibinfo
			{volume} {718-719}},\ \bibinfo {pages} {1} (\bibinfo {year}
		{2017})}\BibitemShut {NoStop}%
	\bibitem [{\citenamefont {Thompson}\ \emph {et~al.}(2008)\citenamefont
		{Thompson}, \citenamefont {Zwickl}, \citenamefont {Jayich}, \citenamefont
		{Marquardt}, \citenamefont {Girvin},\ and\ \citenamefont
		{Harris}}]{Thompson2008}%
	\BibitemOpen
	\bibfield  {author} {\bibinfo {author} {\bibfnamefont {J.~D.}\ \bibnamefont
			{Thompson}}, \bibinfo {author} {\bibfnamefont {B.~M.}\ \bibnamefont
			{Zwickl}}, \bibinfo {author} {\bibfnamefont {A.~M.}\ \bibnamefont {Jayich}},
		\bibinfo {author} {\bibfnamefont {F.}~\bibnamefont {Marquardt}}, \bibinfo
		{author} {\bibfnamefont {S.~M.}\ \bibnamefont {Girvin}}, \ and\ \bibinfo
		{author} {\bibfnamefont {J.~G.~E.}\ \bibnamefont {Harris}},\ }\bibfield
	{title} {\enquote {\bibinfo {title} {Strong dispersive coupling of a
				high-finesse cavity to a micromechanical membrane},}\ }\href {\doibase
		10.1038/nature06715} {\bibfield  {journal} {\bibinfo  {journal} {Nature
				(London)}\ }\textbf {\bibinfo {volume} {452}},\ \bibinfo {pages} {72}
		(\bibinfo {year} {2008})}\BibitemShut {NoStop}%
	\bibitem [{\citenamefont {Bruschi}\ and\ \citenamefont
		{Xuereb}(2018)}]{Bruschi2018}%
	\BibitemOpen
	\bibfield  {author} {\bibinfo {author} {\bibfnamefont {D.~E.}\ \bibnamefont
			{Bruschi}}\ and\ \bibinfo {author} {\bibfnamefont {A.}~\bibnamefont
			{Xuereb}},\ }\bibfield  {title} {\enquote {\bibinfo {title}
			{{M}echano-optics: an optomechanical quantum simulator},}\ }\href {\doibase
		10.1088/1367-2630/aaca27} {\bibfield  {journal} {\bibinfo  {journal} {New J.
				of Phys.}\ }\textbf {\bibinfo {volume} {20}},\ \bibinfo {pages} {065004}
		(\bibinfo {year} {2018})}\BibitemShut {NoStop}%
	\bibitem [{\citenamefont {Aspelmeyer}\ \emph
		{et~al.}(2014{\natexlab{b}})\citenamefont {Aspelmeyer}, \citenamefont
		{Kippenberg},\ and\ \citenamefont {Marquardt}}]{Aspelmeyer14}%
	\BibitemOpen
	\bibfield  {author} {\bibinfo {author} {\bibfnamefont {M.}~\bibnamefont
			{Aspelmeyer}}, \bibinfo {author} {\bibfnamefont {T.~J.}\ \bibnamefont
			{Kippenberg}}, \ and\ \bibinfo {author} {\bibfnamefont {F.}~\bibnamefont
			{Marquardt}},\ }\bibfield  {title} {\enquote {\bibinfo {title} {Cavity
				optomechanics},}\ }\href {\doibase 10.1103/RevModPhys.86.1391} {\bibfield
		{journal} {\bibinfo  {journal} {Rev. Mod. Phys.}\ }\textbf {\bibinfo {volume}
			{86}},\ \bibinfo {pages} {1391} (\bibinfo {year}
		{2014}{\natexlab{b}})}\BibitemShut {NoStop}%
	\bibitem [{\citenamefont {Kockum}\ \emph {et~al.}(2019)\citenamefont {Kockum},
		\citenamefont {Miranowicz}, \citenamefont {Liberato}, \citenamefont
		{Savasta},\ and\ \citenamefont {Nori}}]{FriskKockum2019}%
	\BibitemOpen
	\bibfield  {author} {\bibinfo {author} {\bibfnamefont {A.~F.}\ \bibnamefont
			{Kockum}}, \bibinfo {author} {\bibfnamefont {A.}~\bibnamefont {Miranowicz}},
		\bibinfo {author} {\bibfnamefont {S.~De}\ \bibnamefont {Liberato}}, \bibinfo
		{author} {\bibfnamefont {S.}~\bibnamefont {Savasta}}, \ and\ \bibinfo
		{author} {\bibfnamefont {F.}~\bibnamefont {Nori}},\ }\bibfield  {title}
	{\enquote {\bibinfo {title} {Ultrastrong coupling between light and
				matter},}\ }\href {\doibase 10.1038/s42254-018-0006-2} {\bibfield  {journal}
		{\bibinfo  {journal} {Nature Reviews Physics}\ }\textbf {\bibinfo {volume}
			{1}},\ \bibinfo {pages} {19--40} (\bibinfo {year} {2019})}\BibitemShut
	{NoStop}%
	\bibitem [{\citenamefont {Heikkil\"a}\ \emph {et~al.}(2014)\citenamefont
		{Heikkil\"a}, \citenamefont {Massel}, \citenamefont {Tuorila}, \citenamefont
		{Khan},\ and\ \citenamefont {Sillanp\"a\"a}}]{Heik14}%
	\BibitemOpen
	\bibfield  {author} {\bibinfo {author} {\bibfnamefont {T.~T.}\ \bibnamefont
			{Heikkil\"a}}, \bibinfo {author} {\bibfnamefont {F.}~\bibnamefont {Massel}},
		\bibinfo {author} {\bibfnamefont {J.}~\bibnamefont {Tuorila}}, \bibinfo
		{author} {\bibfnamefont {R.}~\bibnamefont {Khan}}, \ and\ \bibinfo {author}
		{\bibfnamefont {M.~A.}\ \bibnamefont {Sillanp\"a\"a}},\ }\bibfield  {title}
	{\enquote {\bibinfo {title} {Enhancing optomechanical coupling via the
				{J}osephson effect},}\ }\href {\doibase 10.1103/PhysRevLett.112.203603}
	{\bibfield  {journal} {\bibinfo  {journal} {Phys. Rev. Lett.}\ }\textbf
		{\bibinfo {volume} {112}},\ \bibinfo {pages} {203603} (\bibinfo {year}
		{2014})}\BibitemShut {NoStop}%
	\bibitem [{\citenamefont {Pirkkalainen}\ \emph {et~al.}(2015)\citenamefont
		{Pirkkalainen}, \citenamefont {Cho}, \citenamefont {Massel}, \citenamefont
		{Tuorila}, \citenamefont {Heikkil\"{a}}, \citenamefont {Hakonen},\ and\
		\citenamefont {Sillanp\"{a}\"{a}}}]{Pirkkalainen2015}%
	\BibitemOpen
	\bibfield  {author} {\bibinfo {author} {\bibfnamefont {J.-M.}\ \bibnamefont
			{Pirkkalainen}}, \bibinfo {author} {\bibfnamefont {S.U.}\ \bibnamefont
			{Cho}}, \bibinfo {author} {\bibfnamefont {F.}~\bibnamefont {Massel}},
		\bibinfo {author} {\bibfnamefont {J.}~\bibnamefont {Tuorila}}, \bibinfo
		{author} {\bibfnamefont {T.T.}\ \bibnamefont {Heikkil\"{a}}}, \bibinfo
		{author} {\bibfnamefont {P.J.}\ \bibnamefont {Hakonen}}, \ and\ \bibinfo
		{author} {\bibfnamefont {M.A.}\ \bibnamefont {Sillanp\"{a}\"{a}}},\
	}\bibfield  {title} {\enquote {\bibinfo {title} {Cavity optomechanics
				mediated by a quantum two-level system},}\ }\href
	{https://doi.org/10.1038/ncomms7981} {\bibfield  {journal} {\bibinfo
			{journal} {Nat. Commun.}\ }\textbf {\bibinfo {volume} {6}},\ \bibinfo {pages}
		{6981} (\bibinfo {year} {2015})}\BibitemShut {NoStop}%
	\bibitem [{\citenamefont {Zhao}\ \emph {et~al.}(2015)\citenamefont {Zhao},
		\citenamefont {Liu}, \citenamefont {Liu},\ and\ \citenamefont
		{Nori}}]{Zhao15}%
	\BibitemOpen
	\bibfield  {author} {\bibinfo {author} {\bibfnamefont {Y.~J.}\ \bibnamefont
			{Zhao}}, \bibinfo {author} {\bibfnamefont {Y.~L.}\ \bibnamefont {Liu}},
		\bibinfo {author} {\bibfnamefont {Y.~X.}\ \bibnamefont {Liu}}, \ and\
		\bibinfo {author} {\bibfnamefont {F.}~\bibnamefont {Nori}},\ }\bibfield
	{title} {\enquote {\bibinfo {title} {Generating nonclassical photon states
				via longitudinal couplings between superconducting qubits and microwave
				fields},}\ }\href {http://link.aps.org/doi/10.1103/PhysRevA.91.053820}
	{\bibfield  {journal} {\bibinfo  {journal} {Phys. Rev. A}\ }\textbf {\bibinfo
			{volume} {91}},\ \bibinfo {pages} {053820} (\bibinfo {year}
		{2015})}\BibitemShut {NoStop}%
	\bibitem [{\citenamefont {Richer}\ and\ \citenamefont
		{DiVincenzo}(2016)}]{Richer16}%
	\BibitemOpen
	\bibfield  {author} {\bibinfo {author} {\bibfnamefont {S.}~\bibnamefont
			{Richer}}\ and\ \bibinfo {author} {\bibfnamefont {D.}~\bibnamefont
			{DiVincenzo}},\ }\bibfield  {title} {\enquote {\bibinfo {title} {Circuit
				design implementing longitudinal coupling: A scalable scheme for
				superconducting qubits},}\ }\href {\doibase 10.1103/PhysRevB.93.134501}
	{\bibfield  {journal} {\bibinfo  {journal} {Phys. Rev. B}\ }\textbf {\bibinfo
			{volume} {93}},\ \bibinfo {pages} {134501} (\bibinfo {year}
		{2016})}\BibitemShut {NoStop}%
	\bibitem [{\citenamefont {Richer}\ \emph {et~al.}(2017)\citenamefont {Richer},
		\citenamefont {Maleeva}, \citenamefont {Skacel}, \citenamefont {Pop},\ and\
		\citenamefont {DiVincenzo}}]{Richer17}%
	\BibitemOpen
	\bibfield  {author} {\bibinfo {author} {\bibfnamefont {S.}~\bibnamefont
			{Richer}}, \bibinfo {author} {\bibfnamefont {N.}~\bibnamefont {Maleeva}},
		\bibinfo {author} {\bibfnamefont {S.~T.}\ \bibnamefont {Skacel}}, \bibinfo
		{author} {\bibfnamefont {I.~M.}\ \bibnamefont {Pop}}, \ and\ \bibinfo
		{author} {\bibfnamefont {D.}~\bibnamefont {DiVincenzo}},\ }\bibfield  {title}
	{\enquote {\bibinfo {title} {Inductively shunted transmon qubit with tunable
				transverse and longitudinal coupling},}\ }\href {\doibase
		10.1103/PhysRevB.96.174520} {\bibfield  {journal} {\bibinfo  {journal} {Phys.
				Rev. B}\ }\textbf {\bibinfo {volume} {96}},\ \bibinfo {pages} {174520}
		(\bibinfo {year} {2017})}\BibitemShut {NoStop}%
	\bibitem [{\citenamefont {Zueco}\ \emph {et~al.}(2009)\citenamefont {Zueco},
		\citenamefont {Reuther}, \citenamefont {Kohler},\ and\ \citenamefont
		{H\"anggi}}]{Zueco09}%
	\BibitemOpen
	\bibfield  {author} {\bibinfo {author} {\bibfnamefont {D.}~\bibnamefont
			{Zueco}}, \bibinfo {author} {\bibfnamefont {G.~M.}\ \bibnamefont {Reuther}},
		\bibinfo {author} {\bibfnamefont {S.}~\bibnamefont {Kohler}}, \ and\ \bibinfo
		{author} {\bibfnamefont {P.}~\bibnamefont {H\"anggi}},\ }\bibfield  {title}
	{\enquote {\bibinfo {title} {Qubit-oscillator dynamics in the dispersive
				regime: Analytical theory beyond the rotating-wave approximation},}\ }\href
	{\doibase 10.1103/PhysRevA.80.033846} {\bibfield  {journal} {\bibinfo
			{journal} {Phys. Rev. A}\ }\textbf {\bibinfo {volume} {80}},\ \bibinfo
		{pages} {033846} (\bibinfo {year} {2009})}\BibitemShut {NoStop}%
	\bibitem [{\citenamefont {Didier}\ \emph {et~al.}(2015)\citenamefont {Didier},
		\citenamefont {Bourassa},\ and\ \citenamefont {Blais}}]{Didier15}%
	\BibitemOpen
	\bibfield  {author} {\bibinfo {author} {\bibfnamefont {N.}~\bibnamefont
			{Didier}}, \bibinfo {author} {\bibfnamefont {J.}~\bibnamefont {Bourassa}}, \
		and\ \bibinfo {author} {\bibfnamefont {A.}~\bibnamefont {Blais}},\ }\bibfield
	{title} {\enquote {\bibinfo {title} {Fast quantum nondemolition readout by
				parametric modulation of longitudinal qubit-oscillator interaction},}\ }\href
	{http://link.aps.org/doi/10.1103/PhysRevLett.115.203601} {\bibfield
		{journal} {\bibinfo  {journal} {Phys. Rev. Lett.}\ }\textbf {\bibinfo
			{volume} {115}},\ \bibinfo {pages} {203601} (\bibinfo {year}
		{2015})}\BibitemShut {NoStop}%
	\bibitem [{\citenamefont {Cirio}\ \emph {et~al.}(2017)\citenamefont {Cirio},
		\citenamefont {Debnath}, \citenamefont {Lambert},\ and\ \citenamefont
		{Nori}}]{Cirio17}%
	\BibitemOpen
	\bibfield  {author} {\bibinfo {author} {\bibfnamefont {M.}~\bibnamefont
			{Cirio}}, \bibinfo {author} {\bibfnamefont {K.}~\bibnamefont {Debnath}},
		\bibinfo {author} {\bibfnamefont {N.}~\bibnamefont {Lambert}}, \ and\
		\bibinfo {author} {\bibfnamefont {F.}~\bibnamefont {Nori}},\ }\bibfield
	{title} {\enquote {\bibinfo {title} {Amplified optomechanical transduction of
				virtual radiation pressure},}\ }\href {\doibase
		10.1103/PhysRevLett.119.053601} {\bibfield  {journal} {\bibinfo  {journal}
			{Phys. Rev. Lett.}\ }\textbf {\bibinfo {volume} {119}},\ \bibinfo {pages}
		{053601} (\bibinfo {year} {2017})}\BibitemShut {NoStop}%
	\bibitem [{\citenamefont {Rabl}\ \emph {et~al.}(2009)\citenamefont {Rabl},
		\citenamefont {Cappellaro}, \citenamefont {Dutt}, \citenamefont {Jiang},
		\citenamefont {Maze},\ and\ \citenamefont {Lukin}}]{Rabl2009}%
	\BibitemOpen
	\bibfield  {author} {\bibinfo {author} {\bibfnamefont {P.}~\bibnamefont
			{Rabl}}, \bibinfo {author} {\bibfnamefont {P.}~\bibnamefont {Cappellaro}},
		\bibinfo {author} {\bibfnamefont {M.~V.~Gurudev}\ \bibnamefont {Dutt}},
		\bibinfo {author} {\bibfnamefont {L.}~\bibnamefont {Jiang}}, \bibinfo
		{author} {\bibfnamefont {J.~R.}\ \bibnamefont {Maze}}, \ and\ \bibinfo
		{author} {\bibfnamefont {M.~D.}\ \bibnamefont {Lukin}},\ }\bibfield  {title}
	{\enquote {\bibinfo {title} {Strong magnetic coupling between an electronic
				spin qubit and a mechanical resonator},}\ }\href {\doibase
		10.1103/PhysRevB.79.041302} {\bibfield  {journal} {\bibinfo  {journal} {Phys.
				Rev. B}\ }\textbf {\bibinfo {volume} {79}},\ \bibinfo {pages} {041302}
		(\bibinfo {year} {2009})}\BibitemShut {NoStop}%
	\bibitem [{\citenamefont {Poot}\ \emph {et~al.}(2010)\citenamefont {Poot},
		\citenamefont {Etaki}, \citenamefont {Mahboob}, \citenamefont {Onomitsu},
		\citenamefont {Yamaguchi}, \citenamefont {Blanter},\ and\ \citenamefont
		{van~der Zant}}]{Poot2010}%
	\BibitemOpen
	\bibfield  {author} {\bibinfo {author} {\bibfnamefont {M.}~\bibnamefont
			{Poot}}, \bibinfo {author} {\bibfnamefont {S.}~\bibnamefont {Etaki}},
		\bibinfo {author} {\bibfnamefont {I.}~\bibnamefont {Mahboob}}, \bibinfo
		{author} {\bibfnamefont {K.}~\bibnamefont {Onomitsu}}, \bibinfo {author}
		{\bibfnamefont {H.}~\bibnamefont {Yamaguchi}}, \bibinfo {author}
		{\bibfnamefont {Ya.~M.}\ \bibnamefont {Blanter}}, \ and\ \bibinfo {author}
		{\bibfnamefont {H.~S.~J.}\ \bibnamefont {van~der Zant}},\ }\bibfield  {title}
	{\enquote {\bibinfo {title} {Tunable backaction of a {DC SQUID} on an
				integrated micromechanical resonator},}\ }\href {\doibase
		10.1103/PhysRevLett.105.207203} {\bibfield  {journal} {\bibinfo  {journal}
			{Phys. Rev. Lett.}\ }\textbf {\bibinfo {volume} {105}},\ \bibinfo {pages}
		{207203} (\bibinfo {year} {2010})}\BibitemShut {NoStop}%
	\bibitem [{\citenamefont {Teufel}\ \emph {et~al.}(2011)\citenamefont {Teufel},
		\citenamefont {Donner}, \citenamefont {Li}, \citenamefont {Harlow},
		\citenamefont {Allman}, \citenamefont {Cicak}, \citenamefont {Sirois},
		\citenamefont {Whittaker}, \citenamefont {Lehnert},\ and\ \citenamefont
		{Simmonds}}]{Teufel11}%
	\BibitemOpen
	\bibfield  {author} {\bibinfo {author} {\bibfnamefont {J.~D.}\ \bibnamefont
			{Teufel}}, \bibinfo {author} {\bibfnamefont {T.}~\bibnamefont {Donner}},
		\bibinfo {author} {\bibfnamefont {D.-L.}\ \bibnamefont {Li}}, \bibinfo
		{author} {\bibfnamefont {J.-W.}\ \bibnamefont {Harlow}}, \bibinfo {author}
		{\bibfnamefont {M.~S.}\ \bibnamefont {Allman}}, \bibinfo {author}
		{\bibfnamefont {K.}~\bibnamefont {Cicak}}, \bibinfo {author} {\bibfnamefont
			{A.~J.}\ \bibnamefont {Sirois}}, \bibinfo {author} {\bibfnamefont {J.~D.}\
			\bibnamefont {Whittaker}}, \bibinfo {author} {\bibfnamefont {K.~W.}\
			\bibnamefont {Lehnert}}, \ and\ \bibinfo {author} {\bibfnamefont {R.~W.}\
			\bibnamefont {Simmonds}},\ }\bibfield  {title} {\enquote {\bibinfo {title}
			{Sideband cooling of micromechanical motion to the quantum ground state},}\
	}\href {http://dx.doi.org/10.1038/nature10261} {\bibfield  {journal}
		{\bibinfo  {journal} {Nature (London)}\ }\textbf {\bibinfo {volume} {475}},\
		\bibinfo {pages} {359} (\bibinfo {year} {2011})}\BibitemShut {NoStop}%
	\bibitem [{\citenamefont {Viennot}\ \emph {et~al.}(2018)\citenamefont
		{Viennot}, \citenamefont {Ma},\ and\ \citenamefont {Lehnert}}]{Viennot2018}%
	\BibitemOpen
	\bibfield  {author} {\bibinfo {author} {\bibfnamefont {J.~J.}\ \bibnamefont
			{Viennot}}, \bibinfo {author} {\bibfnamefont {X.}~\bibnamefont {Ma}}, \ and\
		\bibinfo {author} {\bibfnamefont {K.~W.}\ \bibnamefont {Lehnert}},\
	}\bibfield  {title} {\enquote {\bibinfo {title} {Phonon-number-sensitive
				electromechanics},}\ }\href {\doibase 10.1103/PhysRevLett.121.183601}
	{\bibfield  {journal} {\bibinfo  {journal} {Phys. Rev. Lett.}\ }\textbf
		{\bibinfo {volume} {121}},\ \bibinfo {pages} {183601} (\bibinfo {year}
		{2018})}\BibitemShut {NoStop}%
	\bibitem [{\citenamefont {Manenti}\ \emph {et~al.}(2016)\citenamefont
		{Manenti}, \citenamefont {Peterer}, \citenamefont {Nersisyan}, \citenamefont
		{Magnusson}, \citenamefont {Patterson},\ and\ \citenamefont
		{Leek}}]{Manenti16}%
	\BibitemOpen
	\bibfield  {author} {\bibinfo {author} {\bibfnamefont {R.}~\bibnamefont
			{Manenti}}, \bibinfo {author} {\bibfnamefont {M.~J.}\ \bibnamefont
			{Peterer}}, \bibinfo {author} {\bibfnamefont {A.}~\bibnamefont {Nersisyan}},
		\bibinfo {author} {\bibfnamefont {E.~B.}\ \bibnamefont {Magnusson}}, \bibinfo
		{author} {\bibfnamefont {A.}~\bibnamefont {Patterson}}, \ and\ \bibinfo
		{author} {\bibfnamefont {P.~J.}\ \bibnamefont {Leek}},\ }\bibfield  {title}
	{\enquote {\bibinfo {title} {Surface acoustic wave resonators in the quantum
				regime},}\ }\href {\doibase 10.1103/PhysRevB.93.041411} {\bibfield  {journal}
		{\bibinfo  {journal} {Phys. Rev. B}\ }\textbf {\bibinfo {volume} {93}},\
		\bibinfo {pages} {041411} (\bibinfo {year} {2016})}\BibitemShut {NoStop}%
	\bibitem [{\citenamefont {Manenti}\ \emph {et~al.}(2017)\citenamefont
		{Manenti}, \citenamefont {Kockum}, \citenamefont {Patterson}, \citenamefont
		{Behrle}, \citenamefont {Rahamim}, \citenamefont {Tancredi}, \citenamefont
		{Nori},\ and\ \citenamefont {Leek}}]{Manenti2017}%
	\BibitemOpen
	\bibfield  {author} {\bibinfo {author} {\bibfnamefont {R.}~\bibnamefont
			{Manenti}}, \bibinfo {author} {\bibfnamefont {A.~F.}\ \bibnamefont {Kockum}},
		\bibinfo {author} {\bibfnamefont {A.}~\bibnamefont {Patterson}}, \bibinfo
		{author} {\bibfnamefont {T.}~\bibnamefont {Behrle}}, \bibinfo {author}
		{\bibfnamefont {J.}~\bibnamefont {Rahamim}}, \bibinfo {author} {\bibfnamefont
			{G.}~\bibnamefont {Tancredi}}, \bibinfo {author} {\bibfnamefont
			{F.}~\bibnamefont {Nori}}, \ and\ \bibinfo {author} {\bibfnamefont {P.~J.}\
			\bibnamefont {Leek}},\ }\bibfield  {title} {\enquote {\bibinfo {title}
			{Circuit quantum acoustodynamics with surface acoustic waves},}\ }\href
	{https://doi.org/10.1038/s41467-017-01063-9} {\bibfield  {journal} {\bibinfo
			{journal} {Nat. Commun.}\ }\textbf {\bibinfo {volume} {8}} (\bibinfo {year}
		{2017})}\BibitemShut {NoStop}%
	\bibitem [{\citenamefont {Gustafsson}\ \emph {et~al.}(2014)\citenamefont
		{Gustafsson}, \citenamefont {Aref}, \citenamefont {Kockum}, \citenamefont
		{Ekstrom}, \citenamefont {Johansson},\ and\ \citenamefont
		{Delsing}}]{Gustafsson2014}%
	\BibitemOpen
	\bibfield  {author} {\bibinfo {author} {\bibfnamefont {M.~V.}\ \bibnamefont
			{Gustafsson}}, \bibinfo {author} {\bibfnamefont {T.}~\bibnamefont {Aref}},
		\bibinfo {author} {\bibfnamefont {A.~F.}\ \bibnamefont {Kockum}}, \bibinfo
		{author} {\bibfnamefont {M.~K.}\ \bibnamefont {Ekstrom}}, \bibinfo {author}
		{\bibfnamefont {G.}~\bibnamefont {Johansson}}, \ and\ \bibinfo {author}
		{\bibfnamefont {P.}~\bibnamefont {Delsing}},\ }\bibfield  {title} {\enquote
		{\bibinfo {title} {Propagating phonons coupled to an artificial atom},}\
	}\href {\doibase 10.1126/science.1257219} {\bibfield  {journal} {\bibinfo
			{journal} {Science}\ }\textbf {\bibinfo {volume} {346}},\ \bibinfo {pages}
		{207} (\bibinfo {year} {2014})}\BibitemShut {NoStop}%
	\bibitem [{\citenamefont {et. al.}(2019)}]{Delsing2019}%
	\BibitemOpen
	\bibfield  {author} {\bibinfo {author} {\bibfnamefont {P.~Delsing}\
			\bibnamefont {et. al.}},\ }\bibfield  {title} {\enquote {\bibinfo {title}
			{The 2019 surface acoustic waves roadmap},}\ }\href {\doibase
		10.1088/1361-6463/ab1b04} {\bibfield  {journal} {\bibinfo  {journal} {J.
				Phys. D}\ }\textbf {\bibinfo {volume} {52}},\ \bibinfo {pages} {353001}
		(\bibinfo {year} {2019})}\BibitemShut {NoStop}%
	\bibitem [{\citenamefont {Kockum}\ \emph {et~al.}(2018)\citenamefont {Kockum},
		\citenamefont {Johansson},\ and\ \citenamefont {Nori}}]{Kockum2018}%
	\BibitemOpen
	\bibfield  {author} {\bibinfo {author} {\bibfnamefont {A.~F.}\ \bibnamefont
			{Kockum}}, \bibinfo {author} {\bibfnamefont {G.}~\bibnamefont {Johansson}}, \
		and\ \bibinfo {author} {\bibfnamefont {F.}~\bibnamefont {Nori}},\ }\bibfield
	{title} {\enquote {\bibinfo {title} {Decoherence-free interaction between
				giant atoms in waveguide quantum electrodynamics},}\ }\href {\doibase
		10.1103/PhysRevLett.120.140404} {\bibfield  {journal} {\bibinfo  {journal}
			{Phys. Rev. Lett.}\ }\textbf {\bibinfo {volume} {120}},\ \bibinfo {pages}
		{140404} (\bibinfo {year} {2018})}\BibitemShut {NoStop}%
	\bibitem [{\citenamefont {Schuetz}\ \emph {et~al.}(2015)\citenamefont
		{Schuetz}, \citenamefont {Kessler}, \citenamefont {Giedke}, \citenamefont
		{Vandersypen}, \citenamefont {Lukin},\ and\ \citenamefont
		{Cirac}}]{Schuetz15}%
	\BibitemOpen
	\bibfield  {author} {\bibinfo {author} {\bibfnamefont {M.~J.~A.}\
			\bibnamefont {Schuetz}}, \bibinfo {author} {\bibfnamefont {E.~M.}\
			\bibnamefont {Kessler}}, \bibinfo {author} {\bibfnamefont {G.}~\bibnamefont
			{Giedke}}, \bibinfo {author} {\bibfnamefont {L.~M.~K.}\ \bibnamefont
			{Vandersypen}}, \bibinfo {author} {\bibfnamefont {M.~D.}\ \bibnamefont
			{Lukin}}, \ and\ \bibinfo {author} {\bibfnamefont {J.~I.}\ \bibnamefont
			{Cirac}},\ }\bibfield  {title} {\enquote {\bibinfo {title} {Universal quantum
				transducers based on surface acoustic waves},}\ }\href {\doibase
		10.1103/PhysRevX.5.031031} {\bibfield  {journal} {\bibinfo  {journal} {Phys.
				Rev. X}\ }\textbf {\bibinfo {volume} {5}},\ \bibinfo {pages} {031031}
		(\bibinfo {year} {2015})}\BibitemShut {NoStop}%
	\bibitem [{\citenamefont {Bolgar}\ \emph {et~al.}(2018)\citenamefont {Bolgar},
		\citenamefont {Zotova}, \citenamefont {Kirichenko}, \citenamefont {Besedin},
		\citenamefont {Semenov}, \citenamefont {Shaikhaidarov},\ and\ \citenamefont
		{Astafiev}}]{Bolgar18}%
	\BibitemOpen
	\bibfield  {author} {\bibinfo {author} {\bibfnamefont {A.~N.}\ \bibnamefont
			{Bolgar}}, \bibinfo {author} {\bibfnamefont {J.~I.}\ \bibnamefont {Zotova}},
		\bibinfo {author} {\bibfnamefont {D.~D.}\ \bibnamefont {Kirichenko}},
		\bibinfo {author} {\bibfnamefont {I.~S.}\ \bibnamefont {Besedin}}, \bibinfo
		{author} {\bibfnamefont {A.~V.}\ \bibnamefont {Semenov}}, \bibinfo {author}
		{\bibfnamefont {R.~S.}\ \bibnamefont {Shaikhaidarov}}, \ and\ \bibinfo
		{author} {\bibfnamefont {O.~V.}\ \bibnamefont {Astafiev}},\ }\bibfield
	{title} {\enquote {\bibinfo {title} {Quantum regime of a two-dimensional
				phonon cavity},}\ }\href {\doibase 10.1103/PhysRevLett.120.223603} {\bibfield
		{journal} {\bibinfo  {journal} {Phys. Rev. Lett.}\ }\textbf {\bibinfo
			{volume} {120}},\ \bibinfo {pages} {223603} (\bibinfo {year}
		{2018})}\BibitemShut {NoStop}%
	\bibitem [{\citenamefont {You}\ and\ \citenamefont {Nori}(2003)}]{You2003}%
	\BibitemOpen
	\bibfield  {author} {\bibinfo {author} {\bibfnamefont {J.~Q.}\ \bibnamefont
			{You}}\ and\ \bibinfo {author} {\bibfnamefont {F.}~\bibnamefont {Nori}},\
	}\bibfield  {title} {\enquote {\bibinfo {title} {Quantum information
				processing with superconducting qubits in a microwave field},}\ }\href
	{\doibase 10.1103/PhysRevB.68.064509} {\bibfield  {journal} {\bibinfo
			{journal} {Phys. Rev. B}\ }\textbf {\bibinfo {volume} {68}},\ \bibinfo
		{pages} {064509} (\bibinfo {year} {2003})}\BibitemShut {NoStop}%
	\bibitem [{\citenamefont {Irish}\ and\ \citenamefont {Schwab}(2003)}]{Irish03}%
	\BibitemOpen
	\bibfield  {author} {\bibinfo {author} {\bibfnamefont {E.~K.}\ \bibnamefont
			{Irish}}\ and\ \bibinfo {author} {\bibfnamefont {K.}~\bibnamefont {Schwab}},\
	}\bibfield  {title} {\enquote {\bibinfo {title} {Quantum measurement of a
				coupled nanomechanical resonator--{C}ooper-pair box system},}\ }\href
	{\doibase 10.1103/PhysRevB.68.155311} {\bibfield  {journal} {\bibinfo
			{journal} {Phys. Rev. B}\ }\textbf {\bibinfo {volume} {68}},\ \bibinfo
		{pages} {155311} (\bibinfo {year} {2003})}\BibitemShut {NoStop}%
	\bibitem [{\citenamefont {Sun}\ \emph {et~al.}(2006)\citenamefont {Sun},
		\citenamefont {Wei}, \citenamefont {Liu},\ and\ \citenamefont
		{Nori}}]{Sun06}%
	\BibitemOpen
	\bibfield  {author} {\bibinfo {author} {\bibfnamefont {C.~P.}\ \bibnamefont
			{Sun}}, \bibinfo {author} {\bibfnamefont {L.~F.}\ \bibnamefont {Wei}},
		\bibinfo {author} {\bibfnamefont {Yu-xi}\ \bibnamefont {Liu}}, \ and\
		\bibinfo {author} {\bibfnamefont {F.}~\bibnamefont {Nori}},\ }\bibfield
	{title} {\enquote {\bibinfo {title} {Quantum transducers: Integrating
				transmission lines and nanomechanical resonators via charge qubits},}\ }\href
	{\doibase 10.1103/PhysRevA.73.022318} {\bibfield  {journal} {\bibinfo
			{journal} {Phys. Rev. A}\ }\textbf {\bibinfo {volume} {73}},\ \bibinfo
		{pages} {022318} (\bibinfo {year} {2006})}\BibitemShut {NoStop}%
	\bibitem [{\citenamefont {Sete}\ \emph {et~al.}(2015)\citenamefont {Sete},
		\citenamefont {Martinis},\ and\ \citenamefont {Korotkov}}]{Sete2015}%
	\BibitemOpen
	\bibfield  {author} {\bibinfo {author} {\bibfnamefont {E.~A.}\ \bibnamefont
			{Sete}}, \bibinfo {author} {\bibfnamefont {J.~M.}\ \bibnamefont {Martinis}},
		\ and\ \bibinfo {author} {\bibfnamefont {A.~N.}\ \bibnamefont {Korotkov}},\
	}\bibfield  {title} {\enquote {\bibinfo {title} {Quantum theory of a bandpass
				{P}urcell filter for qubit readout},}\ }\href {\doibase
		10.1103/PhysRevA.92.012325} {\bibfield  {journal} {\bibinfo  {journal} {Phys.
				Rev. A}\ }\textbf {\bibinfo {volume} {92}},\ \bibinfo {pages} {012325}
		(\bibinfo {year} {2015})}\BibitemShut {NoStop}%
	\bibitem [{\citenamefont {S\'anchez-Burillo}\ \emph {et~al.}(2016)\citenamefont
		{S\'anchez-Burillo}, \citenamefont {Mart\'{\i}n-Moreno}, \citenamefont
		{Garc\'{\i}a-Ripoll},\ and\ \citenamefont {Zueco}}]{Burillo16}%
	\BibitemOpen
	\bibfield  {author} {\bibinfo {author} {\bibfnamefont {E.}~\bibnamefont
			{S\'anchez-Burillo}}, \bibinfo {author} {\bibfnamefont {L.}~\bibnamefont
			{Mart\'{\i}n-Moreno}}, \bibinfo {author} {\bibfnamefont {J.~J.}\ \bibnamefont
			{Garc\'{\i}a-Ripoll}}, \ and\ \bibinfo {author} {\bibfnamefont
			{D.}~\bibnamefont {Zueco}},\ }\bibfield  {title} {\enquote {\bibinfo {title}
			{Full two-photon down-conversion of a single photon},}\ }\href {\doibase
		10.1103/PhysRevA.94.053814} {\bibfield  {journal} {\bibinfo  {journal} {Phys.
				Rev. A}\ }\textbf {\bibinfo {volume} {94}},\ \bibinfo {pages} {053814}
		(\bibinfo {year} {2016})}\BibitemShut {NoStop}%
	\bibitem [{\citenamefont {Johansson}\ \emph
		{et~al.}(2013{\natexlab{b}})\citenamefont {Johansson}, \citenamefont
		{Nation},\ and\ \citenamefont {Nori}}]{Johansson13qutip}%
	\BibitemOpen
	\bibfield  {author} {\bibinfo {author} {\bibfnamefont {J.~R.}\ \bibnamefont
			{Johansson}}, \bibinfo {author} {\bibfnamefont {P.~D.}\ \bibnamefont
			{Nation}}, \ and\ \bibinfo {author} {\bibfnamefont {F.}~\bibnamefont
			{Nori}},\ }\bibfield  {title} {\enquote {\bibinfo {title} {Qutip 2: {A}
				{P}ython framework for the dynamics of open quantum systems},}\ }\href
	{http://www.sciencedirect.com/science/article/pii/S0010465512003955}
	{\bibfield  {journal} {\bibinfo  {journal} {Comput. Phys. Commun.}\ }\textbf
		{\bibinfo {volume} {184}},\ \bibinfo {pages} {1234} (\bibinfo {year}
		{2013}{\natexlab{b}})}\BibitemShut {NoStop}%
	\bibitem [{\citenamefont {Johansson}\ \emph {et~al.}(2012)\citenamefont
		{Johansson}, \citenamefont {Nation},\ and\ \citenamefont
		{Nori}}]{Johansson12qutip}%
	\BibitemOpen
	\bibfield  {author} {\bibinfo {author} {\bibfnamefont {J.~R.}\ \bibnamefont
			{Johansson}}, \bibinfo {author} {\bibfnamefont {P.~D.}\ \bibnamefont
			{Nation}}, \ and\ \bibinfo {author} {\bibfnamefont {F.}~\bibnamefont
			{Nori}},\ }\bibfield  {title} {\enquote {\bibinfo {title} {Qutip: {A}n
				open-source {P}ython framework for the dynamics of open quantum systems},}\
	}\href {http://www.sciencedirect.com/science/article/pii/S0010465512000835}
	{\bibfield  {journal} {\bibinfo  {journal} {Comput. Phys. Commun.}\ }\textbf
		{\bibinfo {volume} {183}},\ \bibinfo {pages} {1760} (\bibinfo {year}
		{2012})}\BibitemShut {NoStop}%
	\bibitem [{\citenamefont {Sankey}\ \emph {et~al.}(2010)\citenamefont {Sankey},
		\citenamefont {Yang}, \citenamefont {Zwickl}, \citenamefont {Jayich},\ and\
		\citenamefont {Harris}}]{Sankey2010}%
	\BibitemOpen
	\bibfield  {author} {\bibinfo {author} {\bibfnamefont {J.~C.}\ \bibnamefont
			{Sankey}}, \bibinfo {author} {\bibfnamefont {C.}~\bibnamefont {Yang}},
		\bibinfo {author} {\bibfnamefont {B.~M.}\ \bibnamefont {Zwickl}}, \bibinfo
		{author} {\bibfnamefont {A.~M.}\ \bibnamefont {Jayich}}, \ and\ \bibinfo
		{author} {\bibfnamefont {J.~G.~E.}\ \bibnamefont {Harris}},\ }\bibfield
	{title} {\enquote {\bibinfo {title} {Strong and tunable nonlinear
				optomechanical coupling in a low-loss system},}\ }\href {\doibase
		10.1038/nphys1707} {\bibfield  {journal} {\bibinfo  {journal} {Nat. Phys.}\
		}\textbf {\bibinfo {volume} {6}},\ \bibinfo {pages} {707} (\bibinfo {year}
		{2010})}\BibitemShut {NoStop}%
	\bibitem [{\citenamefont {Liao}\ and\ \citenamefont {Nori}(2014)}]{Liao2014}%
	\BibitemOpen
	\bibfield  {author} {\bibinfo {author} {\bibfnamefont {J.-Q.}\ \bibnamefont
			{Liao}}\ and\ \bibinfo {author} {\bibfnamefont {F.}~\bibnamefont {Nori}},\
	}\bibfield  {title} {\enquote {\bibinfo {title} {Single-photon quadratic
				optomechanics},}\ }\href {https://doi.org/10.1038/srep06302} {\bibfield
		{journal} {\bibinfo  {journal} {Sci. Rep.}\ }\textbf {\bibinfo {volume}
			{4}},\ \bibinfo {pages} {6302} (\bibinfo {year} {2014})}\BibitemShut
	{NoStop}%
	\bibitem [{\citenamefont {Nunnenkamp}\ \emph {et~al.}(2010)\citenamefont
		{Nunnenkamp}, \citenamefont {B\o{}rkje}, \citenamefont {Harris},\ and\
		\citenamefont {Girvin}}]{Nunnenkamp10}%
	\BibitemOpen
	\bibfield  {author} {\bibinfo {author} {\bibfnamefont {A.}~\bibnamefont
			{Nunnenkamp}}, \bibinfo {author} {\bibfnamefont {K.}~\bibnamefont
			{B\o{}rkje}}, \bibinfo {author} {\bibfnamefont {J.~G.~E.}\ \bibnamefont
			{Harris}}, \ and\ \bibinfo {author} {\bibfnamefont {S.~M.}\ \bibnamefont
			{Girvin}},\ }\bibfield  {title} {\enquote {\bibinfo {title} {Cooling and
				squeezing via quadratic optomechanical coupling},}\ }\href {\doibase
		10.1103/PhysRevA.82.021806} {\bibfield  {journal} {\bibinfo  {journal} {Phys.
				Rev. A}\ }\textbf {\bibinfo {volume} {82}},\ \bibinfo {pages} {021806}
		(\bibinfo {year} {2010})}\BibitemShut {NoStop}%
	\bibitem [{\citenamefont {Tan}\ \emph {et~al.}(2013)\citenamefont {Tan},
		\citenamefont {Bariani}, \citenamefont {Li},\ and\ \citenamefont
		{Meystre}}]{Tan13}%
	\BibitemOpen
	\bibfield  {author} {\bibinfo {author} {\bibfnamefont {H.-T.}\ \bibnamefont
			{Tan}}, \bibinfo {author} {\bibfnamefont {F.}~\bibnamefont {Bariani}},
		\bibinfo {author} {\bibfnamefont {G.-X.}\ \bibnamefont {Li}}, \ and\ \bibinfo
		{author} {\bibfnamefont {P.}~\bibnamefont {Meystre}},\ }\bibfield  {title}
	{\enquote {\bibinfo {title} {Generation of macroscopic quantum superpositions
				of optomechanical oscillators by dissipation},}\ }\href {\doibase
		10.1103/PhysRevA.88.023817} {\bibfield  {journal} {\bibinfo  {journal} {Phys.
				Rev. A}\ }\textbf {\bibinfo {volume} {88}},\ \bibinfo {pages} {023817}
		(\bibinfo {year} {2013})}\BibitemShut {NoStop}%
	\bibitem [{\citenamefont {Liao}\ and\ \citenamefont {Nori}(2013)}]{Liao2013}%
	\BibitemOpen
	\bibfield  {author} {\bibinfo {author} {\bibfnamefont {J.-Q.}\ \bibnamefont
			{Liao}}\ and\ \bibinfo {author} {\bibfnamefont {F.}~\bibnamefont {Nori}},\
	}\bibfield  {title} {\enquote {\bibinfo {title} {Photon blockade in
				quadratically coupled optomechanical systems},}\ }\href {\doibase
		10.1103/PhysRevA.88.023853} {\bibfield  {journal} {\bibinfo  {journal} {Phys.
				Rev. A}\ }\textbf {\bibinfo {volume} {88}},\ \bibinfo {pages} {023853}
		(\bibinfo {year} {2013})}\BibitemShut {NoStop}%
	\bibitem [{\citenamefont {Caves}(1982)}]{Caves1982}%
	\BibitemOpen
	\bibfield  {author} {\bibinfo {author} {\bibfnamefont {C.~M.}\ \bibnamefont
			{Caves}},\ }\bibfield  {title} {\enquote {\bibinfo {title} {Quantum limits on
				noise in linear amplifiers},}\ }\href {\doibase 10.1103/PhysRevD.26.1817}
	{\bibfield  {journal} {\bibinfo  {journal} {Phys. Rev. D}\ }\textbf {\bibinfo
			{volume} {26}},\ \bibinfo {pages} {1817} (\bibinfo {year}
		{1982})}\BibitemShut {NoStop}%
	\bibitem [{\citenamefont {Clerk}\ \emph {et~al.}(2010)\citenamefont {Clerk},
		\citenamefont {Devoret}, \citenamefont {Girvin}, \citenamefont {Marquardt},\
		and\ \citenamefont {Schoelkopf}}]{Clerk10}%
	\BibitemOpen
	\bibfield  {author} {\bibinfo {author} {\bibfnamefont {A.~A.}\ \bibnamefont
			{Clerk}}, \bibinfo {author} {\bibfnamefont {M.~H.}\ \bibnamefont {Devoret}},
		\bibinfo {author} {\bibfnamefont {S.~M.}\ \bibnamefont {Girvin}}, \bibinfo
		{author} {\bibfnamefont {Florian}\ \bibnamefont {Marquardt}}, \ and\ \bibinfo
		{author} {\bibfnamefont {R.~J.}\ \bibnamefont {Schoelkopf}},\ }\bibfield
	{title} {\enquote {\bibinfo {title} {Introduction to quantum noise,
				measurement, and amplification},}\ }\href {\doibase
		10.1103/RevModPhys.82.1155} {\bibfield  {journal} {\bibinfo  {journal} {Rev.
				Mod. Phys.}\ }\textbf {\bibinfo {volume} {82}},\ \bibinfo {pages} {1155}
		(\bibinfo {year} {2010})}\BibitemShut {NoStop}%
	\bibitem [{\citenamefont {Blais}\ \emph {et~al.}(2004)\citenamefont {Blais},
		\citenamefont {Huang}, \citenamefont {Wallraff}, \citenamefont {Girvin},\
		and\ \citenamefont {Schoelkopf}}]{Blais04}%
	\BibitemOpen
	\bibfield  {author} {\bibinfo {author} {\bibfnamefont {A.}~\bibnamefont
			{Blais}}, \bibinfo {author} {\bibfnamefont {R.-S.}\ \bibnamefont {Huang}},
		\bibinfo {author} {\bibfnamefont {A.}~\bibnamefont {Wallraff}}, \bibinfo
		{author} {\bibfnamefont {S.~M.}\ \bibnamefont {Girvin}}, \ and\ \bibinfo
		{author} {\bibfnamefont {R.~J.}\ \bibnamefont {Schoelkopf}},\ }\bibfield
	{title} {\enquote {\bibinfo {title} {Cavity quantum electrodynamics for
				superconducting electrical circuits: An architecture for quantum
				computation},}\ }\href {\doibase 10.1103/PhysRevA.69.062320} {\bibfield
		{journal} {\bibinfo  {journal} {Phys. Rev. A}\ }\textbf {\bibinfo {volume}
			{69}},\ \bibinfo {pages} {062320} (\bibinfo {year} {2004})}\BibitemShut
	{NoStop}%
	\bibitem [{\citenamefont {Boissonneault}\ \emph {et~al.}(2008)\citenamefont
		{Boissonneault}, \citenamefont {Gambetta},\ and\ \citenamefont
		{Blais}}]{Boissonneault08}%
	\BibitemOpen
	\bibfield  {author} {\bibinfo {author} {\bibfnamefont {M.}~\bibnamefont
			{Boissonneault}}, \bibinfo {author} {\bibfnamefont {J.~M.}\ \bibnamefont
			{Gambetta}}, \ and\ \bibinfo {author} {\bibfnamefont {A.}~\bibnamefont
			{Blais}},\ }\bibfield  {title} {\enquote {\bibinfo {title} {Nonlinear
				dispersive regime of cavity {QED}: The dressed dephasing model},}\ }\href
	{\doibase 10.1103/PhysRevA.77.060305} {\bibfield  {journal} {\bibinfo
			{journal} {Phys. Rev. A}\ }\textbf {\bibinfo {volume} {77}},\ \bibinfo
		{pages} {060305} (\bibinfo {year} {2008})}\BibitemShut {NoStop}%
	\bibitem [{\citenamefont {Boissonneault}\ \emph {et~al.}(2009)\citenamefont
		{Boissonneault}, \citenamefont {Gambetta},\ and\ \citenamefont
		{Blais}}]{Boissonneault09}%
	\BibitemOpen
	\bibfield  {author} {\bibinfo {author} {\bibfnamefont {M.}~\bibnamefont
			{Boissonneault}}, \bibinfo {author} {\bibfnamefont {J.~M.}\ \bibnamefont
			{Gambetta}}, \ and\ \bibinfo {author} {\bibfnamefont {A.}~\bibnamefont
			{Blais}},\ }\bibfield  {title} {\enquote {\bibinfo {title} {Dispersive regime
				of circuit qed: Photon-dependent qubit dephasing and relaxation rates},}\
	}\href {\doibase 10.1103/PhysRevA.79.013819} {\bibfield  {journal} {\bibinfo
			{journal} {Phys. Rev. A}\ }\textbf {\bibinfo {volume} {79}},\ \bibinfo
		{pages} {013819} (\bibinfo {year} {2009})}\BibitemShut {NoStop}%
	\bibitem [{\citenamefont {Jaskula}\ \emph {et~al.}(2012)\citenamefont
		{Jaskula}, \citenamefont {Partridge}, \citenamefont {Bonneau}, \citenamefont
		{Lopes}, \citenamefont {Ruaudel}, \citenamefont {Boiron},\ and\ \citenamefont
		{Westbrook}}]{Jaskula12}%
	\BibitemOpen
	\bibfield  {author} {\bibinfo {author} {\bibfnamefont {J.-C.}\ \bibnamefont
			{Jaskula}}, \bibinfo {author} {\bibfnamefont {G.~B.}\ \bibnamefont
			{Partridge}}, \bibinfo {author} {\bibfnamefont {M.}~\bibnamefont {Bonneau}},
		\bibinfo {author} {\bibfnamefont {R.}~\bibnamefont {Lopes}}, \bibinfo
		{author} {\bibfnamefont {J.}~\bibnamefont {Ruaudel}}, \bibinfo {author}
		{\bibfnamefont {D.}~\bibnamefont {Boiron}}, \ and\ \bibinfo {author}
		{\bibfnamefont {C.~I.}\ \bibnamefont {Westbrook}},\ }\bibfield  {title}
	{\enquote {\bibinfo {title} {Acoustic analog to the dynamical {C}asimir
				effect in a {B}ose-{E}instein condensate},}\ }\href {\doibase
		10.1103/PhysRevLett.109.220401} {\bibfield  {journal} {\bibinfo  {journal}
			{Phys. Rev. Lett.}\ }\textbf {\bibinfo {volume} {109}},\ \bibinfo {pages}
		{220401} (\bibinfo {year} {2012})}\BibitemShut {NoStop}%
	\bibitem [{\citenamefont {Motazedifard}\ \emph {et~al.}(2017)\citenamefont
		{Motazedifard}, \citenamefont {Naderi},\ and\ \citenamefont
		{Roknizadeh}}]{Motazedifard2017}%
	\BibitemOpen
	\bibfield  {author} {\bibinfo {author} {\bibfnamefont {A.}~\bibnamefont
			{Motazedifard}}, \bibinfo {author} {\bibfnamefont {M.~H.}\ \bibnamefont
			{Naderi}}, \ and\ \bibinfo {author} {\bibfnamefont {R.}~\bibnamefont
			{Roknizadeh}},\ }\bibfield  {title} {\enquote {\bibinfo {title} {Dynamical
				{C}asimir effect of phonon excitation in the dispersive regime of cavity
				optomechanics},}\ }\href {\doibase 10.1364/josab.34.000642} {\bibfield
		{journal} {\bibinfo  {journal} {J. Opt. Soc. Am. B}\ }\textbf {\bibinfo
			{volume} {34}},\ \bibinfo {pages} {642} (\bibinfo {year} {2017})}\BibitemShut
	{NoStop}%
	\bibitem [{\citenamefont {Johansson}\ \emph {et~al.}(2010)\citenamefont
		{Johansson}, \citenamefont {Johansson}, \citenamefont {Wilson},\ and\
		\citenamefont {Nori}}]{Johansson09}%
	\BibitemOpen
	\bibfield  {author} {\bibinfo {author} {\bibfnamefont {J.~R.}\ \bibnamefont
			{Johansson}}, \bibinfo {author} {\bibfnamefont {G.}~\bibnamefont
			{Johansson}}, \bibinfo {author} {\bibfnamefont {C.~M.}\ \bibnamefont
			{Wilson}}, \ and\ \bibinfo {author} {\bibfnamefont {F.}~\bibnamefont
			{Nori}},\ }\bibfield  {title} {\enquote {\bibinfo {title} {Dynamical
				{C}asimir effect in superconducting microwave circuits},}\ }\href {\doibase
		10.1103/PhysRevA.82.052509} {\bibfield  {journal} {\bibinfo  {journal} {Phys.
				Rev. A}\ }\textbf {\bibinfo {volume} {82}},\ \bibinfo {pages} {052509}
		(\bibinfo {year} {2010})}\BibitemShut {NoStop}%
	\bibitem [{\citenamefont {Liu}\ \emph {et~al.}(2010)\citenamefont {Liu},
		\citenamefont {Miranowicz}, \citenamefont {Gao}, \citenamefont {Bajer},
		\citenamefont {Sun},\ and\ \citenamefont {Nori}}]{Liu10}%
	\BibitemOpen
	\bibfield  {author} {\bibinfo {author} {\bibfnamefont {Y.-X.}\ \bibnamefont
			{Liu}}, \bibinfo {author} {\bibfnamefont {A.}~\bibnamefont {Miranowicz}},
		\bibinfo {author} {\bibfnamefont {Y.~B.}\ \bibnamefont {Gao}}, \bibinfo
		{author} {\bibfnamefont {J.}~\bibnamefont {Bajer}}, \bibinfo {author}
		{\bibfnamefont {C.~P.}\ \bibnamefont {Sun}}, \ and\ \bibinfo {author}
		{\bibfnamefont {F.}~\bibnamefont {Nori}},\ }\bibfield  {title} {\enquote
		{\bibinfo {title} {{Qubit-induced phonon blockade as a signature of quantum
					behavior in nanomechanical resonators}},}\ }\href
	{http://link.aps.org/doi/10.1103/PhysRevA.82.032101} {\bibfield  {journal}
		{\bibinfo  {journal} {Phys. Rev. A}\ }\textbf {\bibinfo {volume} {82}},\
		\bibinfo {pages} {032101} (\bibinfo {year} {2010})}\BibitemShut {NoStop}%
	\bibitem [{\citenamefont {Scully}\ and\ \citenamefont
		{Zubairy}(1997)}]{Scully1997}%
	\BibitemOpen
	\bibfield  {author} {\bibinfo {author} {\bibfnamefont {M.~O.}\ \bibnamefont
			{Scully}}\ and\ \bibinfo {author} {\bibfnamefont {M.~S.}\ \bibnamefont
			{Zubairy}},\ }\href@noop {} {\emph {\bibinfo {title} {Quantum {O}ptics}}}\
	(\bibinfo  {publisher} {Cambridge University Press, Cambridge},\ \bibinfo
	{year} {1997})\BibitemShut {NoStop}%
	\bibitem [{\citenamefont {L\"{a}hteenm\"{a}ki}\ \emph
		{et~al.}(2013)\citenamefont {L\"{a}hteenm\"{a}ki}, \citenamefont {Paraoanu},
		\citenamefont {Hassel},\ and\ \citenamefont {Hakonen}}]{Lhteenmki2013}%
	\BibitemOpen
	\bibfield  {author} {\bibinfo {author} {\bibfnamefont {P.}~\bibnamefont
			{L\"{a}hteenm\"{a}ki}}, \bibinfo {author} {\bibfnamefont {G.~S.}\
			\bibnamefont {Paraoanu}}, \bibinfo {author} {\bibfnamefont {J.}~\bibnamefont
			{Hassel}}, \ and\ \bibinfo {author} {\bibfnamefont {P.~J.}\ \bibnamefont
			{Hakonen}},\ }\bibfield  {title} {\enquote {\bibinfo {title} {Dynamical
				{C}asimir effect in a {J}osephson metamaterial},}\ }\href {\doibase
		10.1073/pnas.1212705110} {\bibfield  {journal} {\bibinfo  {journal} {PNAS}\
		}\textbf {\bibinfo {volume} {110}},\ \bibinfo {pages} {4234} (\bibinfo {year}
		{2013})}\BibitemShut {NoStop}%
	\bibitem [{\citenamefont {Lambrecht}\ \emph {et~al.}(1996)\citenamefont
		{Lambrecht}, \citenamefont {Jaekel},\ and\ \citenamefont
		{Reynaud}}]{Lambrecht96}%
	\BibitemOpen
	\bibfield  {author} {\bibinfo {author} {\bibfnamefont {A.}~\bibnamefont
			{Lambrecht}}, \bibinfo {author} {\bibfnamefont {M.}~\bibnamefont {Jaekel}}, \
		and\ \bibinfo {author} {\bibfnamefont {S.}~\bibnamefont {Reynaud}},\
	}\bibfield  {title} {\enquote {\bibinfo {title} {Motion induced radiation
				from a vibrating cavity},}\ }\href {\doibase 10.1103/PhysRevLett.77.615}
	{\bibfield  {journal} {\bibinfo  {journal} {Phys. Rev. Lett.}\ }\textbf
		{\bibinfo {volume} {77}},\ \bibinfo {pages} {615} (\bibinfo {year}
		{1996})}\BibitemShut {NoStop}%
	\bibitem [{\citenamefont {Satzinger}\ \emph {et~al.}(2018)\citenamefont
		{Satzinger}, \citenamefont {Zhong}, \citenamefont {Chang}, \citenamefont
		{Peairs}, \citenamefont {Bienfait}, \citenamefont {Chou}, \citenamefont
		{Cleland}, \citenamefont {Conner}, \citenamefont {Dumur}, \citenamefont
		{Grebel}, \citenamefont {Gutierrez}, \citenamefont {November}, \citenamefont
		{Povey}, \citenamefont {Whiteley}, \citenamefont {Awschalom}, \citenamefont
		{Schuster},\ and\ \citenamefont {Cleland}}]{Satzinger2018}%
	\BibitemOpen
	\bibfield  {author} {\bibinfo {author} {\bibfnamefont {K.~J.}\ \bibnamefont
			{Satzinger}}, \bibinfo {author} {\bibfnamefont {Y.~P.}\ \bibnamefont
			{Zhong}}, \bibinfo {author} {\bibfnamefont {H.~S.}\ \bibnamefont {Chang}},
		\bibinfo {author} {\bibfnamefont {G.~A.}\ \bibnamefont {Peairs}}, \bibinfo
		{author} {\bibfnamefont {A.}~\bibnamefont {Bienfait}}, \bibinfo {author}
		{\bibfnamefont {Ming-Han}\ \bibnamefont {Chou}}, \bibinfo {author}
		{\bibfnamefont {A.~Y.}\ \bibnamefont {Cleland}}, \bibinfo {author}
		{\bibfnamefont {C.~R.}\ \bibnamefont {Conner}}, \bibinfo {author}
		{\bibfnamefont {É}~\bibnamefont {Dumur}}, \bibinfo {author} {\bibfnamefont
			{J.}~\bibnamefont {Grebel}}, \bibinfo {author} {\bibfnamefont
			{I.}~\bibnamefont {Gutierrez}}, \bibinfo {author} {\bibfnamefont {B.~H.}\
			\bibnamefont {November}}, \bibinfo {author} {\bibfnamefont {R.~G.}\
			\bibnamefont {Povey}}, \bibinfo {author} {\bibfnamefont {S.~J.}\ \bibnamefont
			{Whiteley}}, \bibinfo {author} {\bibfnamefont {D.~D.}\ \bibnamefont
			{Awschalom}}, \bibinfo {author} {\bibfnamefont {D.~I.}\ \bibnamefont
			{Schuster}}, \ and\ \bibinfo {author} {\bibfnamefont {A.~N.}\ \bibnamefont
			{Cleland}},\ }\bibfield  {title} {\enquote {\bibinfo {title} {Quantum control
				of surface acoustic-wave phonons},}\ }\href {\doibase
		10.1038/s41586-018-0719-5} {\bibfield  {journal} {\bibinfo  {journal} {Nature
				(London)}\ }\textbf {\bibinfo {volume} {563}},\ \bibinfo {pages} {661--665}
		(\bibinfo {year} {2018})}\BibitemShut {NoStop}%
	\bibitem [{\citenamefont {Kockum}\ \emph {et~al.}(2014)\citenamefont {Kockum},
		\citenamefont {Delsing},\ and\ \citenamefont {Johansson}}]{Kockum2014}%
	\BibitemOpen
	\bibfield  {author} {\bibinfo {author} {\bibfnamefont {A.~F.}\ \bibnamefont
			{Kockum}}, \bibinfo {author} {\bibfnamefont {P.}~\bibnamefont {Delsing}}, \
		and\ \bibinfo {author} {\bibfnamefont {G.}~\bibnamefont {Johansson}},\
	}\bibfield  {title} {\enquote {\bibinfo {title} {Designing
				frequency-dependent relaxation rates and lamb shifts for a giant artificial
				atom},}\ }\href {\doibase 10.1103/PhysRevA.90.013837} {\bibfield  {journal}
		{\bibinfo  {journal} {Phys. Rev. A}\ }\textbf {\bibinfo {volume} {90}},\
		\bibinfo {pages} {013837} (\bibinfo {year} {2014})}\BibitemShut {NoStop}%
	\bibitem [{\citenamefont {Liu}\ \emph {et~al.}(2014)\citenamefont {Liu},
		\citenamefont {Yang}, \citenamefont {Sun},\ and\ \citenamefont
		{Wang}}]{Liu2014}%
	\BibitemOpen
	\bibfield  {author} {\bibinfo {author} {\bibfnamefont {Y.-X.}\ \bibnamefont
			{Liu}}, \bibinfo {author} {\bibfnamefont {C.-X.}\ \bibnamefont {Yang}},
		\bibinfo {author} {\bibfnamefont {H.-C.}\ \bibnamefont {Sun}}, \ and\
		\bibinfo {author} {\bibfnamefont {X.-B.}\ \bibnamefont {Wang}},\ }\bibfield
	{title} {\enquote {\bibinfo {title} {Coexistence of single- and multi-photon
				processes due to longitudinal couplings between superconducting flux qubits
				and external fields},}\ }\href
	{http://stacks.iop.org/1367-2630/16/i=1/a=015031} {\bibfield  {journal}
		{\bibinfo  {journal} {New J. Phys.}\ }\textbf {\bibinfo {volume} {16}},\
		\bibinfo {pages} {015031} (\bibinfo {year} {2014})}\BibitemShut {NoStop}%
	\bibitem [{\citenamefont {Ding}\ \emph {et~al.}(2017)\citenamefont {Ding},
		\citenamefont {Maslennikov}, \citenamefont {Habl\"utzel},\ and\ \citenamefont
		{Matsukevich}}]{Ding17}%
	\BibitemOpen
	\bibfield  {author} {\bibinfo {author} {\bibfnamefont {S.-Q.}\ \bibnamefont
			{Ding}}, \bibinfo {author} {\bibfnamefont {G.}~\bibnamefont {Maslennikov}},
		\bibinfo {author} {\bibfnamefont {R.}~\bibnamefont {Habl\"utzel}}, \ and\
		\bibinfo {author} {\bibfnamefont {D.}~\bibnamefont {Matsukevich}},\
	}\bibfield  {title} {\enquote {\bibinfo {title} {Cross-{K}err nonlinearity
				for phonon counting},}\ }\href {\doibase 10.1103/PhysRevLett.119.193602}
	{\bibfield  {journal} {\bibinfo  {journal} {Phys. Rev. Lett.}\ }\textbf
		{\bibinfo {volume} {119}},\ \bibinfo {pages} {193602} (\bibinfo {year}
		{2017})}\BibitemShut {NoStop}%
	\bibitem [{\citenamefont {Munro}\ \emph {et~al.}(2005)\citenamefont {Munro},
		\citenamefont {Nemoto}, \citenamefont {Beausoleil},\ and\ \citenamefont
		{Spiller}}]{Munro05}%
	\BibitemOpen
	\bibfield  {author} {\bibinfo {author} {\bibfnamefont {W.~J.}\ \bibnamefont
			{Munro}}, \bibinfo {author} {\bibfnamefont {K.}~\bibnamefont {Nemoto}},
		\bibinfo {author} {\bibfnamefont {R.~G.}\ \bibnamefont {Beausoleil}}, \ and\
		\bibinfo {author} {\bibfnamefont {T.~P.}\ \bibnamefont {Spiller}},\
	}\bibfield  {title} {\enquote {\bibinfo {title} {High-efficiency
				quantum-nondemolition single-photon-number-resolving detector},}\ }\href
	{\doibase 10.1103/PhysRevA.71.033819} {\bibfield  {journal} {\bibinfo
			{journal} {Phys. Rev. A}\ }\textbf {\bibinfo {volume} {71}},\ \bibinfo
		{pages} {033819} (\bibinfo {year} {2005})}\BibitemShut {NoStop}%
\end{thebibliography}
\end{document}